\documentclass[fleqn,usenatbib]{mnras}

\usepackage[T1]{fontenc}
\DeclareRobustCommand{\VAN}[3]{#2}
\let\VANthebibliography\thebibliography
\def\thebibliography{\DeclareRobustCommand{\VAN}[3]{##3}\VANthebibliography}
\usepackage{graphicx}
\usepackage{amsmath}
\usepackage{amssymb}
\usepackage{physics}
\usepackage{longtable}
\usepackage{orcidlink}
\usepackage{multirow}
\usepackage{anyfontsize}
\title[Electron-Impact Excitation of Zr~{\sc i}-{\sc iii}]{Electron-Impact Excitation of Zirconium~{\sc i}-{\sc iii} in support of Neutron Star Merger Diagnostics}
\author[M. McCann, C. P. Ballance, F. McNeill, S. A. Sim, and C. A. Ramsbottom]{
M. McCann$^{1}$\thanks{E-mail: michael.mccann@qub.ac.uk}\orcidlink{0000-0002-1532-1240}, C. P. Ballance$^{1}$\orcidlink{0000-0003-1693-1793}, F. McNeill$^{1}$\orcidlink{0009-0001-9528-7475}, S. A. Sim$^{1}$\orcidlink{0000-0002-9774-1192}, and C. A. Ramsbottom$^{1}$\orcidlink{0000-0003-1579-8556} \\
$^{1}$Astrophysics Research Centre, School of Mathematics and Physics, Queen's University Belfast, BT7 1NN, Northern Ireland\\
}

\date{Accepted XXX. Received YYY; in original form ZZZ}
\pubyear{2023}

\begin{document}
\label{firstpage}
\pagerange{\pageref{firstpage}--\pageref{lastpage}}
\maketitle

\begin{abstract}
Recent observation and analysis of kilonovae (KNe) spectra as a result of neutron star mergers require accurate and complete atomic structure and collisional data for interpretation. Ideally, the atomic datasets for elements predicted to be abundant in the ejecta 
should be experimentally calibrated. For near-neutral ion stages of Zirconium in particular, the 
A-values and the associated 
excitation/de-excitation rates are required from collision calculations built upon accurate structure models. The atomic orbitals required to perform the structure calculations may be calculated using a Multi-Configuration-Dirac-Fock (MCDF) approximation implemented within the General Relativistic Atomic Structure Package ({\sc grasp0}). Optimized sets of relativistic atomic orbitals are then imported into electron-impact excitation collision calculations. A relativistic $R$-matrix formulation within the Dirac Atomic R-matrix Code ({\sc darc}) is employed to compute collision strengths, which are subsequently Maxwellian convolved to produce excitation/de-excitation rates for a wide range of electron temperatures. These atomic datasets subsequently provide the foundations for non-local thermodynamic equilibrium (NLTE) collisional-radiative models. In this work all these computations have been carried out for the first three ion stages of Zirconium (Zr {\sc i}-{\sc iii}) with the data further interfaced with collisional-radiative and radiative transfer codes to produce synthetic spectra which can be compared with observation.
\end{abstract}


\section{Introduction}\label{introsection}

Binary neutron star mergers (NSM) have been proposed as one possible origin of heavy elements with atomic number greater than iron. 
In the conditions of NSM these elements are expected to be formed through the process of rapid neutron capture (r-process) \citep{kajino2019}, a process which occurs when the timescale for neutron capture is shorter than the timescale for $\beta$ decay. In NSM the comparatively large neutron flux drives the neutron capture time down. The r-process produces a large amount of unstable nuclei which can subsequently decay to form elements that may account for observed abundances in the universe. There has been a significant increase in activity surrounding these topics since the detection of gravitational waves from a NSM event in 2017 and subsequent observations of the produced kilonova (KNe) AT2017gfo \citep{pian2017,smartt2017}.

To determine if heavy species are produced in a NSM it is important to analyse the observed spectra for lines associated with high $Z$ elements. However, disentangling these lines in the observed KNe spectra is a difficult endeavour. To accomplish this we require accurate synthetic spectra to compare against the KNe spectra so that we may distinguish the features, emerging from  multiple ion stages, that belong to the elements of interest. This is further complicated as the KNe plasma rapidly enters a Non-Local Thermodynamic Equilibrium (NLTE) regime at the temperatures and densities of interest, as discussed in \cite{Pognan2023}. To generate a synthetic spectra for a species in a NLTE environment, accurate and complete atomic data sets are necessary, data sets which include energy levels, transition rates and excitation/de-excitation rates for all transitions among the levels of interest. In \cite{Vieira_2023} Zr is suggested as one of the elements contributing to the ejecta spectrum and in \cite{Gillanders_2024} one of the proposed species of interest is Zr {\sc ii}. This prediction was based on 1D radiative transfer simulations using the {\sc tardis} modelling code \citep{tardis}. To aid in the continued analysis of NSM this work will introduce new atomic data for the first three ion stages Zr {\sc i}, Zr {\sc ii} and Zr {\sc iii} into such radiative transfer simulations.

Recent papers by \cite{ljung_2006} and \cite{Lawler2022} have reported experimental energy levels for Zr {\sc i} and Zr {\sc ii} and oscillator strengths for transitions among these levels for Zr {\sc ii}. Additional calculations by \cite{Biemont1981} determined oscillator strengths for transitions in Zr {\sc i} and Zr {\sc ii}. These works are supplemented by data listed in the National Institute of Science and Technology Atomic Spectra Database (NIST ASD) \citep{nist}, where atomic energy level data for Zr {\sc i} and Zr {\sc ii} from \cite{Moore_1971} are reported. Experimental energy levels and oscillator strengths for Zr {\sc iii} are found in the early work of \cite{Reader_1997}. The experimental data from these publications will be used to benchmark the atomic data computed in this work. Our primary goal is comprehensive coverage of the first three ion stages of Zr that provides a reasonable representation of the atomic structure and electron-impact excitation rates to model plasmas at KNe densities and temperatures. It should be noted that there are no collisional excitation data currently available in the literature for any of the three Zr species under consideration. Additionally it is considered that excitation and dielectronic recombination may be the most relevant processes that effect the observed spectra, therefore generating the excitation data will be useful in understanding the KNe.

The generation of the atomic data is split into two sections, first an atomic structure model is constructed and secondly this model is incorporated into a collision calculation to compute the electron-impact excitation cross sections and associated parameters. The atomic structure calculation generates energy levels, oscillator strengths and spontaneous transition probabilities, from a set of orthogonal electron orbitals. The general-purpose relativistic atomic structure program ({\sc grasp0}) \citep{grasp}, which uses the Multi-Configurational Dirac Fock method, is employed in the generation of the electron orbitals. These electron orbitals are then used as the starting point in the second phase, an electron-impact excitation calculation, which generates Maxwellian averaged collision rates over a wide range of electron temperatures which can be used in LTE and NLTE plasma modelling. In this second phase the fully relativistic Dirac Atomic R-matrix Codes ({\sc darc}) \citep{rmatrix,Norrington_1987} are used to compute the collision cross sections, these codes are well-suited to atomic species with high $Z$ where relativistic effects are important. The underlying theory of electron-impact excitation R-matrix calculations is discussed in greater detail within \cite{burke2011} and will not be reproduced here.

The remainder of the paper is structured as follows. In Sec.~\ref{structuresection} the atomic structure models describing the Zr {\sc i}, {\sc ii} and {\sc iii} targets are discussed. The computed energy levels and transition probabilities for transitions among these levels are compared with known experimental values and other theoretical predictions from the literature. In Sec.~\ref{eiesection} results from the corresponding electron-impact excitation calculations are reported, and some examples of both the resulting collision strengths and Maxwellian averaged effective collision strengths for selected allowed and forbidden lines are shown. In Sec.~\ref{modellingsection} the new atomic data that have been generated are incorporated into the collisional radiative modelling code ColRadPy~\citep{colradpy}, to produce synthetic spectra. Analysing these spectra allows us to test the accuracy of the data and probe LTE and NLTE assumptions via the level populations. In Sec.~\ref{sec:tardis} the Zr data is incorporated into the 1D radiative transfer {\sc tardis} code to produce synthetic spectra which are compared to the AT2017gfo observations. Finally in Sec.~\ref{conclusionsection} this work on the atomic data for Zr {\sc i} - {\sc iii} will be summarised and conclusions drawn.

\section{Atomic Structure}\label{structuresection}

The atomic structure models for Zr {\sc i}, {\sc ii} and {\sc iii} have been calculated using an extensively modified version of {\sc grasp0} \citep{grasp}. For Zr {\sc i} the {\sc grasp0} model included a total of 14 non-relativistic configurations, 10 of which were even parity and 4 were odd parity. This configuration set resulted in a total of 726 individual fine-structure energy levels. For Zr {\sc ii} the {\sc grasp0} model included a total of 9 non-relativistic configurations, 6 of which were even parity and 3 were odd parity. This much smaller configuration set resulted in a total of 241 energy levels. Finally, for Zr {\sc iii} the {\sc grasp0} calculation included a total of 28 non-relativistic configurations, 21 of which were even parity and 7 were odd parity. This configuration set resulted in a total of 265 fine-structure energy levels. In Table~\ref{configurationstable} all the configurations included for each ion stage model are listed, noting that only the valence electrons are listed and each configuration includes a Kr core.

\begin{table}
	\centering
	\begin{tabular}{ c c c c c c c }
		\hline
        \multicolumn{2}{c}{Zr {\sc i}} & \multicolumn{2}{c}{Zr {\sc ii}} & \multicolumn{3}{c}{Zr {\sc iii}} \\
        Even & Odd & Even & Odd & \multicolumn{2}{c}{Even} & Odd \\
		\hline
        4d$^2$5s$^2$ & 4d$^2$5s5p & 4d$^2$5s & 4d$^2$5p & 4d$^2$ & 4d6d & 4d4f \\
        4d$^3$5s & 4d5s$^2$5p & 4d$^3$ & 4d5s5p & 4f$^2$ & 6d$^2$ & 4d5p \\
        4d$^4$ & 4d$^3$5p & 4d5s$^2$ & 4d$^2$6p & 4d5s & 6f$^2$ & 4d5f \\
        4d$^2$5p$^2$ & 4d$^2$5s6p & 4d$^2$6s &  & 5s$^2$ & 4d7s & 4d6p \\
        5s$^2$5p$^2$ &  & 4d$^2$5d &  & 5p$^2$ & 7s$^2$ & 4d6f \\
        4d$^2$5s5d &  & 4d5s6s &  & 4d5d & 7p$^2$ & 4d7p \\
        4d5s$^2$5d &  &  &  & 5d$^2$ & 4d7d & 5s5p \\
        5d$^4$ &  &  &  & 5f$^2$ & 7d$^2$ &  \\
        4d$^2$5s6s &  &  &  & 4d6s & 5s5d &  \\
        4d$^2$6s$^2$ &  &  &  & 6s$^2$ & 5s6s &  \\
        &  &  &  & 6p$^2$ & &  \\
        \hline
	\end{tabular}
    \caption{Configurations included in the {\sc grasp0} calculations for each ion stage of Zr, the electron configurations shown list the valence electrons only and each configuration includes a Kr core.}
    \label{configurationstable}
\end{table}

\begin{figure}
    \centering
    \includegraphics[width=\columnwidth]{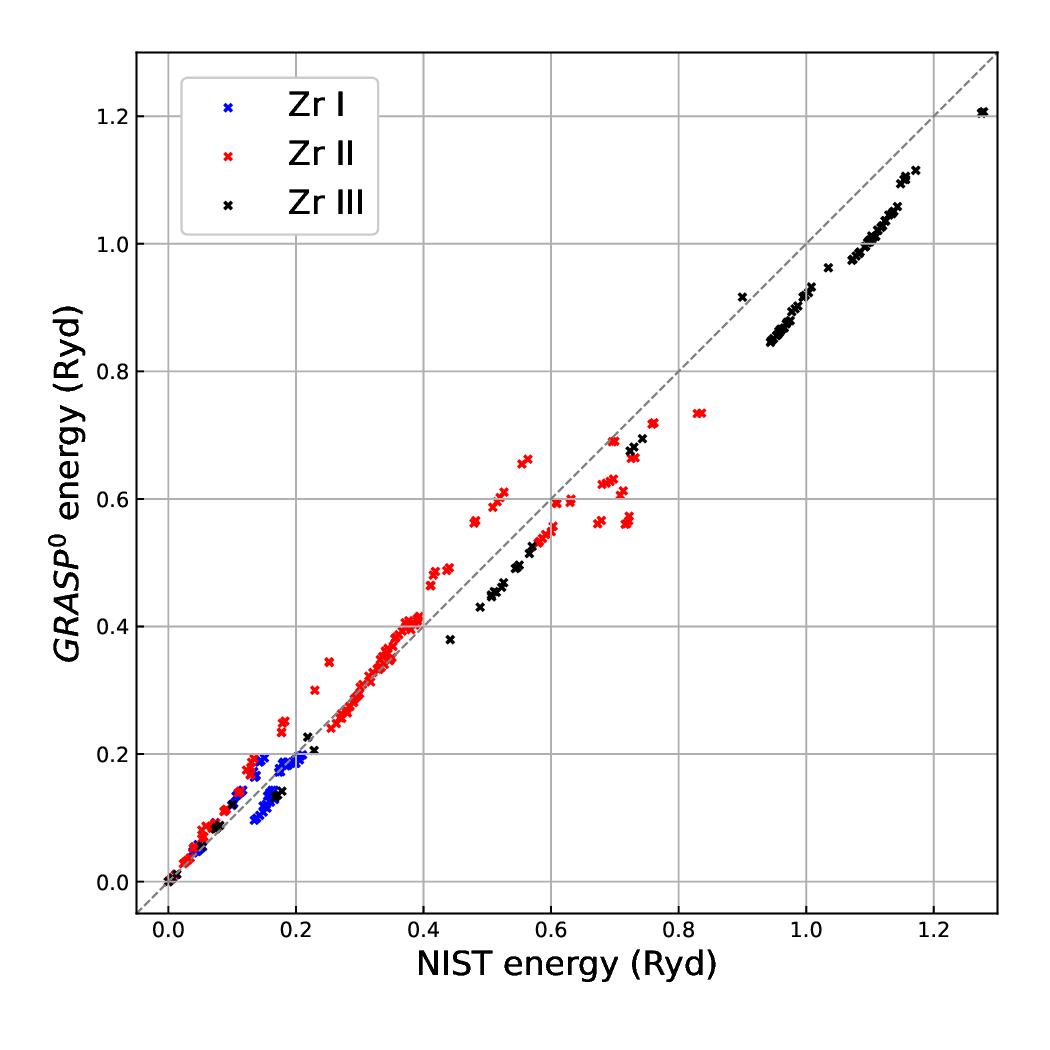}
    \caption{Comparison of {\sc grasp0} energy levels for Zr {\sc i} (blue), Zr {\sc ii} (red) and Zr {\sc iii} (black) with those of \protect\cite{Lawler2022} and \protect\cite{Reader_1997} available in NIST ADS~\protect\citep{nist}. The dotted line represents the line of equality.}
    \label{energylevelsfig}
\end{figure}

A necessary test of  atomic structure accuracy is to compare the absolute energy difference between the calculated level energies from {\sc grasp0} with the experimentally measured values available in the NIST ASD \citep{nist}. In Table~\ref{adf04energylevelszri} we present the energies in Ryds for the first 10 even and first 10 odd levels of Zr {\sc i} and compare with the listed values in \cite{Lawler2022}. For the low-lying even levels the agreement is very good with the largest difference of 0.01~Ryds occurring for the 4d$^{2}$5s$^{2}$ $^{1}$D$_2$ (index 9) level. For the higher lying odd states the agreement consistently falls between $\approx$ 0.02 - 0.04~Ryds which is satisfactory for this neutral case. A similar table of energies is presented in Table~\ref{adf04energylevelszrii} for Zr {\sc ii} where the comparison is made with the listed values in \cite{Lawler2022}. Again good agreement is evident for all the even states listed, differences of at worst 0.013~Ryds are recorded. For this singly ionised system the odd states are much better behaved and agreement of less than 0.01~Ryds is found for all levels presented. Finally, in Table~\ref{adf04energylevelszriii} the comparison is shown for Zr {\sc iii} where the NIST listed values are from \cite{Reader_1997}. The alignment for the even levels is again very satisfactory, the higher lying odd states display disparities between $\approx$ 0.05 - 0.06~Ryds. To graphically display this data we present, in Fig.~\ref{energylevelsfig}, a plot of the full set of available NIST energy levels in Ryds against the {\sc grasp0} energies for each ion species of Zr with the dashed line indicating the line of equality. Clustering around this line for all three charge states is evident.

A further test to gauge the accuracy of the Zr {\sc i}, {\sc ii} and {\sc iii} target state wavefunctions is to compare known values of the Einstein A-values for transitions among the target levels included in each model.
Within {\sc grasp0} there is an option to shift the level energies of the target to their exact experimental positions prior to the computation of the A-values. This is an important feature as the transition probabilities directly depend on the wavelength to an odd power, $\lambda^3$ for dipole transitions (E1 and M1), $\lambda^5$ for quadrupoles (E2 and M2) and $\lambda^7$ for octupoles (E3 and M3). Hence, even small disparities in $\lambda$ can lead to large inaccuracies in the computed A-values. Ensuring spectroscopically accurate values for the wavelengths should significantly improve the accuracy of the A-values and aid the identification of specific lines during the analysis of observational spectra. 
Note that shifting to experimental values may correct the wavelength dependence of an A-value, but not the underlying line strength.
The A-values in this paper are computed using the spectroscopically accurate NIST values listed in Tables~\ref{adf04energylevelszri}-\ref{adf04energylevelszriii}. 
The shifted A-value for dipole transitions can be determined by
\begin{eqnarray}
A_{ij}(\mathrm{Shifted)} = 
   \Big( \frac{\lambda_{\mathrm{Calculated}}}{\lambda_{\mathrm{NIST}}}\Big )^3A_{ij}(\mathrm{Unshifted}) ,
\end{eqnarray}
where the power of 3 would be increased to 5 for quadrupoles and 7 for octupoles.

\begin{figure}
    \centering
    \includegraphics[width=\columnwidth]{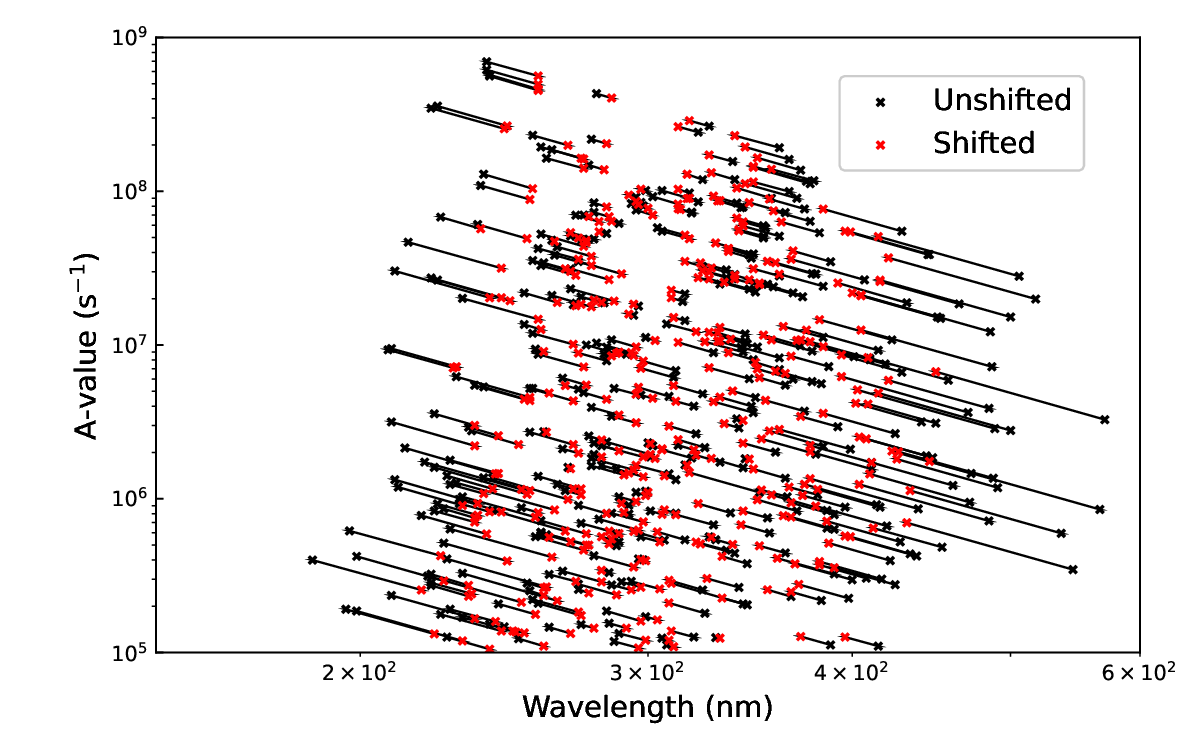}
    \caption{Computed A-values (s$^{-1}$) for a selection of E1 and M1 dipole lines in Zr {\sc ii}, as a function of wavelength (nm) for computations employing a shift to NIST energy values (red) versus those where the {\it ab initio} energies are adopted (black).}
    \label{avalueshift_dipoles}
\end{figure}

\begin{figure}
    \centering
    \includegraphics[width=\columnwidth]{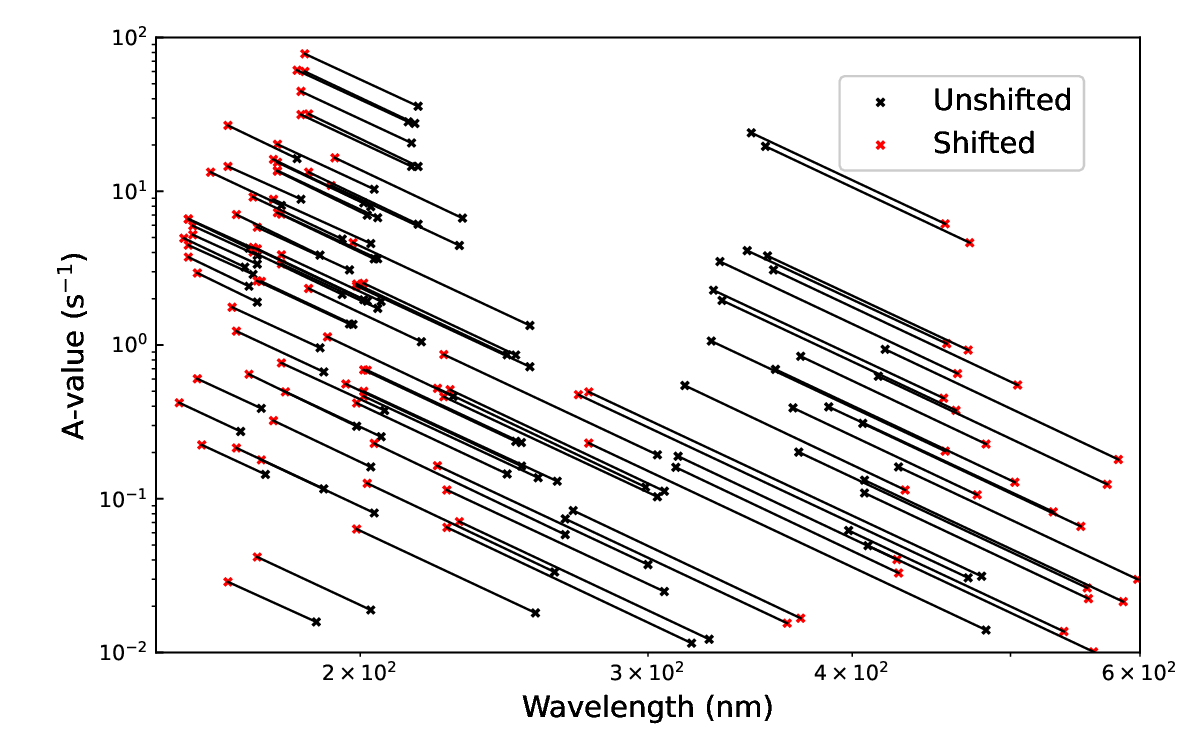}
    \caption{Computed A-values (s$^{-1}$ for a selection of E2 and M2 quadrupole lines in Zr {\sc ii}, as a function of wavelength (nm) for computations employing a shift to NIST energy values (red) versus those where the {\it ab initio} energies are adopted (black).}
    \label{avalueshift_quadrupoles}
\end{figure}

It is important to investigate the effect shifting to experimental energies, where available, has on the resulting A-values. 
In Fig.~\ref{avalueshift_dipoles} a selection of Zr {\sc ii} dipole transitions (both E1 and M1) are presented for wavelengths in the region 150 -- 600 nm, particularly relevant for KNe modelling. Only those transitions among the lowest 100 levels are considered. 
In Fig.~\ref{avalueshift_quadrupoles} a corresponding figure for a selection of quadrupole transitions in Zr {\sc ii} is shown. Clearly the quadrupole lines exhibit a steeper gradient between shifted and non-shifted values as expected from the different power relationship they have with wavelength, with a dependence of $\lambda^{5}$ for quadrupoles rather than $\lambda^{3}$ for dipoles. Additionally the magnitude of the shift gets larger at higher wavelength, 
because the wavelength is inversely proportional to the energy separation $\Delta$E, so an equivalent shift in the energy levels represents a larger proportion of the smaller $\Delta$E at higher wavelength. These two figures emphasise the importance of calibrating with experimental energies, where available, when computing radiative atomic data for use in spectroscopic modelling.

\begin{figure}
    \centering
    \includegraphics[width=\columnwidth]{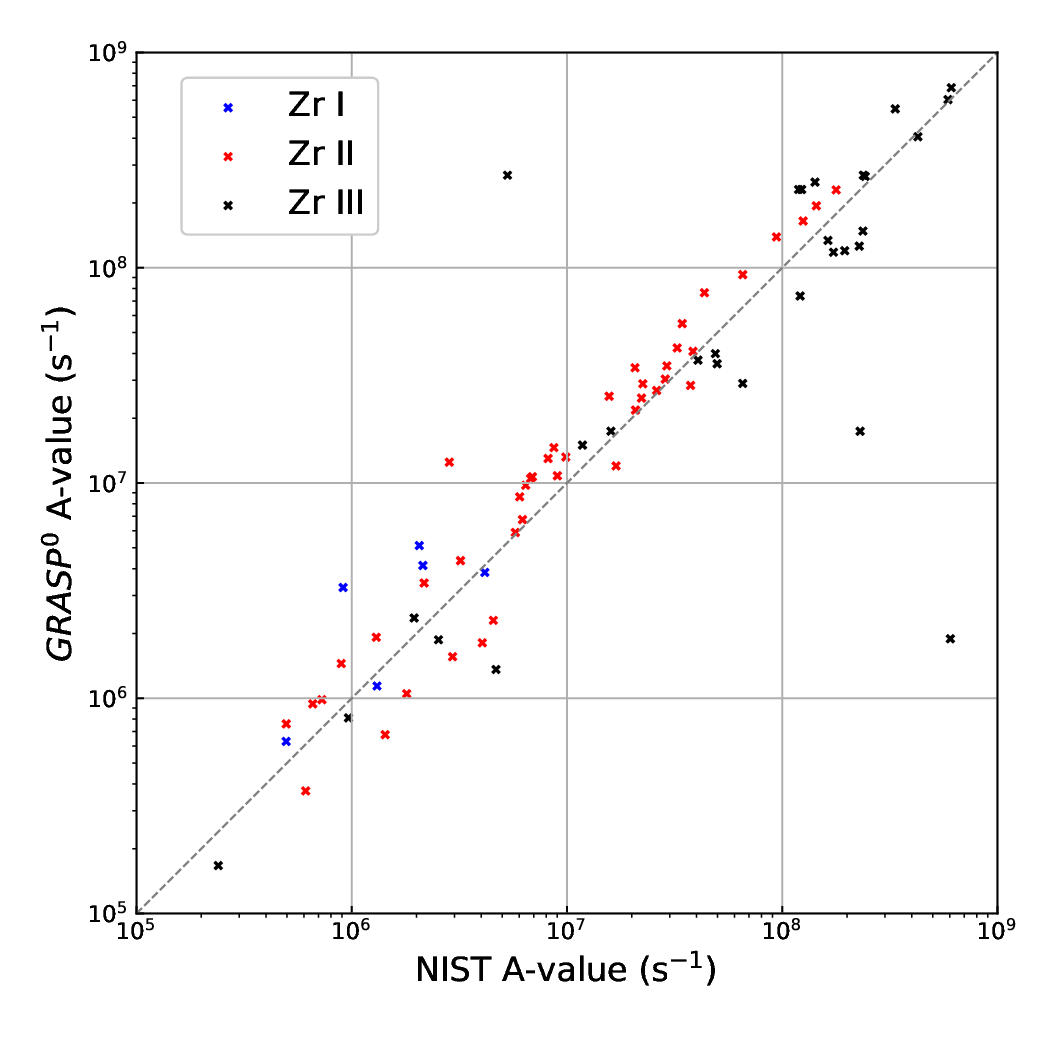}
    \caption{Comparison of {\sc grasp0} A-values for Zr {\sc i} (blue), Zr {\sc ii} (red) and Zr {\sc iii} (black) with those of \protect\cite{Biemont1981,ljung_2006,Reader_1997}.}
    \label{avaluessfig}
\end{figure}

In Table~\ref{zriavaluecomparisontable} the computed A-values for the strongest E1 dipole lines of Zr {\sc i} are compared to the A-values derived from oscillator strengths in~\cite{Biemont1981}. Only a small number of E1 lines are compared here as the target state energies available in NIST only allow for the shifting of 63 of the low lying levels. Good consistency in the order of magnitude between the experimental and calculated values is observed for all transitions considered, the largest disparity observed for the 4-48 dipole line. In Table~\ref{zriiavaluecomparisontable} the calculated A-values for Zr {\sc ii} are compared with the A-values derived from oscillator strengths in~\cite{ljung_2006}. Again only the strong E1 lines are tabulated but there are significantly more to consider as there are many more energies available in NIST for Zr {\sc ii}. Again a similar broad agreement is evident between the available literature and the newly calculated A-values. Finally, in Table~\ref{zriiiavaluecomparisontable} a similar comparison for Zr {\sc iii} is presented. In this case, the comparison is with the A-values listed in \cite{Reader_1997}. Again good agreement is found for the majority of lines. Similar to Fig.~\ref{energylevelsfig} the A-values can be compared graphically, as shown in Fig.~\ref{avaluessfig}. As in Fig.~\ref{energylevelsfig} clustering around the line is evident for all the charge states.

\onecolumn
\begin{longtable}{ c c c c c c c c }
	\hline
    Level & Conf. & Parity & Term & \textit{J} & \cite{Lawler2022} & GRASP$^0$  &  diff \\
          &       &        &      &            & (Ryd)      & (Ryd)      &  (Ryd) \\
	\hline
	1 & 4d$^2$5s$^2$ & even & $^3$F & 2 & 0.0000000 & 0.0000000 & 0.0000000 \\
	2 & 4d$^2$5s$^2$ & even & $^3$F & 3 & 0.0051980 & 0.0047690 & -0.0004290 \\
	3 & 4d$^2$5s$^2$ & even & $^3$F & 4 & 0.0113074 & 0.0106163 & -0.0006911 \\
	4 & 4d$^2$5s$^2$ & even & $^3$P & 2 & 0.0381466 & 0.0471997 & 0.0090531 \\
	5 & 4d$^2$5s$^2$ & even & $^3$P & 0 & 0.0382445 & 0.0448361 & 0.0065916 \\
	6 & 4d$^2$5s$^2$ & even & $^3$P & 1 & 0.0398796 & 0.0462170 & 0.0063374 \\
	7 & 4d$^3$5s$^1$ & even & $^5$F & 1 & 0.0443835 & 0.0471652 & 0.0027817 \\
	8 & 4d$^3$5s$^1$ & even & $^5$F & 2 & 0.0457767 & 0.0482713 & 0.0024946 \\
	9 & 4d$^2$5s$^2$ & even & $^1$D & 2 & 0.0464899 & 0.0580415 & 0.0115516 \\
	10 & 4d$^3$5s$^1$ & even & $^5$F & 3 & 0.0478330 & 0.0499145 & 0.0020815 \\
	26 & 4d$^2$5s$^1$5p$^1$ & odd & $^5$G & 2 & 0.1347175 & 0.0961847 & -0.0385328 \\
	30 & 4d$^2$5s$^1$5p$^1$ & odd & $^5$G & 3 & 0.1385241 & 0.0994675 & -0.0390566 \\
	33 & 4d$^2$5s$^1$5p$^1$ & odd & $^5$G & 4 & 0.1432545 & 0.1037616 & -0.0394928 \\
	37 & 4d$^2$5s$^1$5p$^1$ & odd & $^5$F & 2 & 0.1485047 & 0.1186320 & -0.0298726 \\
	38 & 4d$^2$5s$^1$5p$^1$ & odd & $^5$G & 5 & 0.1486911 & 0.1089937 & -0.0396974 \\
	41 & 4d$^2$5s$^1$5p$^1$ & odd & $^5$F & 1 & 0.1529738 & 0.1168793 & -0.0360945 \\
	42 & 4d$^2$5s$^1$5p$^1$ & odd & $^5$F & 3 & 0.1534932 & 0.1212376 & -0.0322556 \\
	44 & 4d$^2$5s$^1$5p$^1$ & odd & $^5$G & 6 & 0.1547176 & 0.1150933 & -0.0396243 \\
	45 & 4d$^2$5s$^1$5p$^1$ & odd & $^3$F & 2 & 0.1554586 & 0.1340496 & -0.0214090 \\
	46 & 4d$^2$5s$^1$5p$^1$ & odd & $^3$F & 3 & 0.1587625 & 0.1374634 & -0.0212991 \\
	\hline
    \\
    \caption{Comparison between the first 10 even and the first 10 odd energy levels from the GRASP$^0$ calculation of Zr {\sc i} with energies from~\protect\cite{Lawler2022}, in Ryds. The index numbers of the levels are from the energy order of the shifted calculation.}
    \label{adf04energylevelszri}
\end{longtable}

\begin{longtable}{ c c c c c c c c }
	\hline
    Level & Conf. & Parity & Term & \textit{J} & \cite{Lawler2022} & GRASP$^0$  &  diff \\
          &       &        &      &            & (Ryd)      & (Ryd)      & (Ryd)  \\
	\hline
	1 & 4d$^2$5s$^1$ & even & $^4$F & 1.5 & 0.0000000 & 0.0000000 & 0.0000000 \\
	2 & 4d$^2$5s$^1$ & even & $^4$F & 2.5 & 0.0028675 & 0.0027692 & -0.0000983 \\
	3 & 4d$^2$5s$^1$ & even & $^4$F & 3.5 & 0.0069570 & 0.0066571 & -0.0002999 \\
	4 & 4d$^2$5s$^1$ & even & $^4$F & 4.5 & 0.0120552 & 0.0115605 & -0.0004947 \\
	5 & 4d$^3$ & even & $^4$F & 1.5 & 0.0234397 & 0.0281406 & 0.0047009 \\
	6 & 4d$^3$ & even & $^4$F & 2.5 & 0.0263816 & 0.0304560 & 0.0040744 \\
	7 & 4d$^3$ & even & $^4$F & 3.5 & 0.0300685 & 0.0335215 & 0.0034530 \\
	8 & 4d$^3$ & even & $^4$F & 4.5 & 0.0342423 & 0.0372035 & 0.0029612 \\
	9 & 4d$^2$5s$^1$ & even & $^2$D & 1.5 & 0.0387134 & 0.0522989 & 0.0135855 \\
	10 & 4d$^2$5s$^1$ & even & $^2$D & 2.5 & 0.0410571 & 0.0548323 & 0.0137752 \\
	38 & 4d$^2$5p$^1$ & odd & $^4$G & 2.5 & 0.2550074 & 0.2403789 & -0.0146285 \\
	39 & 4d$^2$5p$^1$ & odd & $^4$G & 3.5 & 0.2634386 & 0.2478361 & -0.0156025 \\
	40 & 4d$^2$5p$^1$ & odd & $^2$F & 2.5 & 0.2688691 & 0.2569563 & -0.0119128 \\
	41 & 4d$^2$5p$^1$ & odd & $^2$D & 1.5 & 0.2713535 & 0.2630735 & -0.0082799 \\
	42 & 4d$^2$5p$^1$ & odd & $^4$G & 4.5 & 0.2719209 & 0.2557986 & -0.0161222 \\
	43 & 4d$^2$5p$^1$ & odd & $^4$F & 1.5 & 0.2773476 & 0.2679613 & -0.0093863 \\
	44 & 4d$^2$5p$^1$ & odd & $^4$F & 2.5 & 0.2784056 & 0.2692824 & -0.0091231 \\
	45 & 4d$^2$5p$^1$ & odd & $^2$F & 3.5 & 0.2784992 & 0.2673287 & -0.0111705 \\
	46 & 4d$^2$5p$^1$ & odd & $^4$G & 5.5 & 0.2806314 & 0.2643171 & -0.0163142 \\
	47 & 4d$^2$5p$^1$ & odd & $^2$D & 2.5 & 0.2839512 & 0.2744373 & -0.0095139 \\
	\hline
    \\
    \caption{Comparison between the first 10 even and the first 10 odd energy levels from the GRASP$^0$ calculation of Zr {\sc ii} with energies from~\protect\cite{Lawler2022}, in Ryds. The index numbers of the levels are from the energy order of the shifted calculation.}
    \label{adf04energylevelszrii}
\end{longtable}

\newpage
\begin{longtable}{ c c c c c c c c }
	\hline
    Level & Conf. & Parity & Term & \textit{J} & \cite{Reader_1997} & GRASP$^0$  & diff \\
          &       &        &      &            & (Ryd)      & (Ryd)      & (Ryd) \\
	\hline
	1 & 4d$^2$ & even & $^3$F & 2 & 0.0000000 & 0.0000000 & 0.0000000 \\
	2 & 4d$^2$ & even & $^3$F & 3 & 0.0062111 & 0.0051968 & -0.0010143 \\
	3 & 4d$^2$ & even & $^3$F & 4 & 0.0135455 & 0.0115437 & -0.0020018 \\
	4 & 4d$^2$ & even & $^1$D & 2 & 0.0523376 & 0.0552481 & 0.0029105 \\
	5 & 4d$^2$ & even & $^3$P & 0 & 0.0734812 & 0.0824057 & 0.0089245 \\
	6 & 4d$^2$ & even & $^3$P & 1 & 0.0758823 & 0.0843364 & 0.0084541 \\
	7 & 4d$^2$ & even & $^3$P & 2 & 0.0805557 & 0.0880235 & 0.0074678 \\
	8 & 4d$^2$ & even & $^1$G & 4 & 0.1007002 & 0.1207737 & 0.0200735 \\
	9 & 4d$^1$5s$^1$ & even & $^3$D & 1 & 0.1676838 & 0.1323546 & -0.0353292 \\
	10 & 4d$^1$5s$^1$ & even & $^3$D & 2 & 0.1713665 & 0.1358710 & -0.0354955 \\
	15 & 4d$^1$5p$^1$ & odd & $^1$D & 2 & 0.4888847 & 0.4302977 & -0.0585871 \\
	16 & 4d$^1$5p$^1$ & odd & $^3$F & 2 & 0.5062779 & 0.4494225 & -0.0568555 \\
	17 & 4d$^1$5p$^1$ & odd & $^3$D & 1 & 0.5068091 & 0.4465398 & -0.0602693 \\
	18 & 4d$^1$5p$^1$ & odd & $^3$F & 3 & 0.5110139 & 0.4548248 & -0.0561891 \\
	19 & 4d$^1$5p$^1$ & odd & $^3$D & 2 & 0.5142987 & 0.4534983 & -0.0608005 \\
	20 & 4d$^1$5p$^1$ & odd & $^3$D & 3 & 0.5226011 & 0.4614351 & -0.0611661 \\
	21 & 4d$^1$5p$^1$ & odd & $^3$F & 4 & 0.5256563 & 0.4687638 & -0.0568925 \\
	22 & 4d$^1$5p$^1$ & odd & $^3$P & 1 & 0.5440215 & 0.4908779 & -0.0531436 \\
	23 & 4d$^1$5p$^1$ & odd & $^3$P & 0 & 0.5462827 & 0.4924677 & -0.0538151 \\
	24 & 4d$^1$5p$^1$ & odd & $^3$P & 2 & 0.5500259 & 0.4962671 & -0.0537589 \\
	\hline
    \\
    \caption{Comparison between the first 10 even and the first 10 odd energy levels from the GRASP$^0$ calculation of Zr {\sc iii} with energies from~\protect\cite{Reader_1997}, in Ryds. The index numbers of the levels are from the energy order of the shifted calculation.}
    \label{adf04energylevelszriii}
\end{longtable}

\begin{longtable}{ c c c c c c c c c c c c }
	\hline
	\multicolumn{5}{c}{Lower} & \multicolumn{5}{c}{upper} & \multicolumn{2}{c}{A-values (s$^{-1}$)} \\
	Level & Conf. & Parity & Term & J & Level & Conf. & Parity & Term & J & GRASP$^0$ & \cite{Biemont1981} \\
	\hline
	1  & 4d$^2$5s$^2$ & even & $^3$F & 2 & 45  & 4d$^2$5s$^1$5p$^1$ & odd & $^3$F & 2 & 5.12E+06 & 2.06E+06 \\
	2  & 4d$^2$5s$^2$ & even & $^3$F & 3 & 46  & 4d$^2$5s$^1$5p$^1$ & odd & $^3$F & 3 & 4.14E+06 & 2.14E+06 \\
	4  & 4d$^2$5s$^2$ & even & $^3$P & 2 & 48  & 4d$^2$5s$^1$5p$^1$ & odd & $^3$P & 2 & 3.27E+06 & 9.12E+05 \\
	8  & 4d$^3$5s$^1$ & even & $^5$F & 2 & 50  & 4d$^2$5s$^1$5p$^1$ & odd & $^5$D & 2 & 1.14E+06 & 1.31E+06 \\
	11 & 4d$^3$5s$^1$ & even & $^5$F & 4 & 52  & 4d$^2$5s$^1$5p$^1$ & odd & $^5$D & 3 & 3.84E+06 & 4.15E+06 \\
	11 & 4d$^3$5s$^1$ & even & $^5$F & 4 & 53  & 4d$^2$5s$^1$5p$^1$ & odd & $^5$F & 5 & 6.30E+05 & 4.96E+05 \\
	\hline
    \\
    \caption{Comparison between the A-values from the GRASP$^0$ calculation of Zr {\sc i}, with A-values from~\protect\cite{Biemont1981}. The index numbers of the levels are from the energy order of the shifted calculation.}
    \label{zriavaluecomparisontable}
\end{longtable}
\newpage
\begin{longtable}{ c c c c c c c c c c c c }
	\hline
	\multicolumn{5}{c}{Lower} & \multicolumn{5}{c}{upper} & \multicolumn{2}{c}{A-values (s$^{-1}$)} \\
	Level & Conf. & Parity & Term & J & Level & Conf. & Parity & Term & J & GRASP$^0$ & \cite{ljung_2006} \\
	\hline
	1  & 4d$^2$5s$^1$ & even & $^4$F & 1.5 & 38 & 4d$^2$5p$^1$ & odd & $^4$G & 2.5 & 1.39E+08 & 9.40E+07 \\
	1  & 4d$^2$5s$^1$ & even & $^4$F & 1.5 & 40 & 4d$^2$5p$^1$ & odd & $^2$F & 2.5 & 2.84E+07 & 3.75E+07 \\
	1  & 4d$^2$5s$^1$ & even & $^4$F & 1.5 & 41 & 4d$^2$5p$^1$ & odd & $^2$D & 1.5 & 4.24E+07 & 3.25E+07 \\
	1  & 4d$^2$5s$^1$ & even & $^4$F & 1.5 & 43 & 4d$^2$5p$^1$ & odd & $^4$F & 1.5 & 9.29E+07 & 6.55E+07 \\
	1  & 4d$^2$5s$^1$ & even & $^4$F & 1.5 & 47 & 4d$^2$5p$^1$ & odd & $^2$D & 2.5 & 1.92E+06 & 1.30E+06 \\
	2  & 4d$^2$5s$^1$ & even & $^4$F & 2.5 & 38 & 4d$^2$5p$^1$ & odd & $^4$G & 2.5 & 2.89E+07 & 2.25E+07 \\
	2  & 4d$^2$5s$^1$ & even & $^4$F & 2.5 & 39 & 4d$^2$5p$^1$ & odd & $^4$G & 3.5 & 1.65E+08 & 1.25E+08 \\
	2  & 4d$^2$5s$^1$ & even & $^4$F & 2.5 & 40 & 4d$^2$5p$^1$ & odd & $^2$F & 2.5 & 2.30E+06 & 4.55E+06 \\
	2  & 4d$^2$5s$^1$ & even & $^4$F & 2.5 & 41 & 4d$^2$5p$^1$ & odd & $^2$D & 1.5 & 2.69E+07 & 2.61E+07 \\
	2  & 4d$^2$5s$^1$ & even & $^4$F & 2.5 & 43 & 4d$^2$5p$^1$ & odd & $^4$F & 1.5 & 1.30E+07 & 8.17E+06 \\
	2  & 4d$^2$5s$^1$ & even & $^4$F & 2.5 & 45 & 4d$^2$5p$^1$ & odd & $^2$F & 3.5 & 1.20E+07 & 1.69E+07 \\
	2  & 4d$^2$5s$^1$ & even & $^4$F & 2.5 & 47 & 4d$^2$5p$^1$ & odd & $^2$D & 2.5 & 3.04E+07 & 2.86E+07 \\
	3  & 4d$^2$5s$^1$ & even & $^4$F & 3.5 & 38 & 4d$^2$5p$^1$ & odd & $^4$G & 2.5 & 9.42E+05 & 6.59E+05 \\
	3  & 4d$^2$5s$^1$ & even & $^4$F & 3.5 & 39 & 4d$^2$5p$^1$ & odd & $^4$G & 3.5 & 3.50E+07 & 2.91E+07 \\
	3  & 4d$^2$5s$^1$ & even & $^4$F & 3.5 & 40 & 4d$^2$5p$^1$ & odd & $^2$F & 2.5 & 1.56E+06 & 2.94E+06 \\
	3  & 4d$^2$5s$^1$ & even & $^4$F & 3.5 & 42 & 4d$^2$5p$^1$ & odd & $^4$G & 4.5 & 1.94E+08 & 1.44E+08 \\
	4  & 4d$^2$5s$^1$ & even & $^4$F & 4.5 & 42 & 4d$^2$5p$^1$ & odd & $^4$G & 4.5 & 2.48E+07 & 2.22E+07 \\
	4  & 4d$^2$5s$^1$ & even & $^4$F & 4.5 & 45 & 4d$^2$5p$^1$ & odd & $^2$F & 3.5 & 6.77E+05 & 1.43E+06 \\
	4  & 4d$^2$5s$^1$ & even & $^4$F & 4.5 & 46 & 4d$^2$5p$^1$ & odd & $^4$G & 5.5 & 2.30E+08 & 1.78E+08 \\
	5  &       4d$^3$ & even & $^4$F & 1.5 & 38 & 4d$^2$5p$^1$ & odd & $^4$G & 2.5 & 8.63E+06 & 6.03E+06 \\
	5  &       4d$^3$ & even & $^4$F & 1.5 & 41 & 4d$^2$5p$^1$ & odd & $^2$D & 1.5 & 4.09E+07 & 3.85E+07 \\
	5  &       4d$^3$ & even & $^4$F & 1.5 & 43 & 4d$^2$5p$^1$ & odd & $^4$F & 1.5 & 3.43E+07 & 2.07E+07 \\
	5  &       4d$^3$ & even & $^4$F & 1.5 & 47 & 4d$^2$5p$^1$ & odd & $^2$D & 2.5 & 9.85E+05 & 7.27E+05 \\
	6  &       4d$^3$ & even & $^4$F & 2.5 & 39 & 4d$^2$5p$^1$ & odd & $^4$G & 3.5 & 9.76E+06 & 6.43E+06 \\
	6  &       4d$^3$ & even & $^4$F & 2.5 & 41 & 4d$^2$5p$^1$ & odd & $^2$D & 1.5 & 3.43E+06 & 2.17E+06 \\
	6  &       4d$^3$ & even & $^4$F & 2.5 & 43 & 4d$^2$5p$^1$ & odd & $^4$F & 1.5 & 1.32E+07 & 9.87E+06 \\
	6  &       4d$^3$ & even & $^4$F & 2.5 & 47 & 4d$^2$5p$^1$ & odd & $^2$D & 2.5 & 4.36E+06 & 3.20E+06 \\
	7  &       4d$^3$ & even & $^4$F & 3.5 & 40 & 4d$^2$5p$^1$ & odd & $^2$F & 2.5 & 3.71E+05 & 6.11E+05 \\
	7  &       4d$^3$ & even & $^4$F & 3.5 & 42 & 4d$^2$5p$^1$ & odd & $^4$G & 4.5 & 1.05E+07 & 6.77E+06 \\
	7  &       4d$^3$ & even & $^4$F & 3.5 & 45 & 4d$^2$5p$^1$ & odd & $^2$F & 3.5 & 7.60E+05 & 4.96E+05 \\
	7  &	   4d$^3$ & even & $^4$F & 3.5 & 47 & 4d$^2$5p$^1$ & odd & $^2$D & 2.5 & 6.76E+06 & 6.21E+06 \\
	8  &	   4d$^3$ & even & $^4$F & 4.5 & 45 & 4d$^2$5p$^1$ & odd & $^2$F & 3.5 & 1.05E+06 & 1.80E+06 \\
	8  &	   4d$^3$ & even & $^4$F & 4.5 & 46 & 4d$^2$5p$^1$ & odd & $^4$G & 5.5 & 1.07E+07 & 6.91E+06 \\
	9  & 4d$^2$5s$^1$ & even & $^2$D & 1.5 & 38 & 4d$^2$5p$^1$ & odd & $^4$G & 2.5 & 5.89E+06 & 5.76E+06 \\
	9  & 4d$^2$5s$^1$ & even & $^2$D & 1.5 & 40 & 4d$^2$5p$^1$ & odd & $^2$F & 2.5 & 5.50E+07 & 3.43E+07 \\
	9  & 4d$^2$5s$^1$ & even & $^2$D & 1.5 & 41 & 4d$^2$5p$^1$ & odd & $^2$D & 1.5 & 2.53E+07 & 1.57E+07 \\
	9  & 4d$^2$5s$^1$ & even & $^2$D & 1.5 & 43 & 4d$^2$5p$^1$ & odd & $^4$F & 1.5 & 1.46E+07 & 8.69E+06 \\
	9  & 4d$^2$5s$^1$ & even & $^2$D & 1.5 & 47 & 4d$^2$5p$^1$ & odd & $^2$D & 2.5 & 1.08E+07 & 9.02E+06 \\
	10 & 4d$^2$5s$^1$ & even & $^2$D & 2.5 & 38 & 4d$^2$5p$^1$ & odd & $^4$G & 2.5 & 1.81E+06 & 4.04E+06 \\
	10 & 4d$^2$5s$^1$ & even & $^2$D & 2.5 & 39 & 4d$^2$5p$^1$ & odd & $^4$G & 3.5 & 1.45E+06 & 8.94E+05 \\
	10 & 4d$^2$5s$^1$ & even & $^2$D & 2.5 & 40 & 4d$^2$5p$^1$ & odd & $^2$F & 2.5 & 2.18E+07 & 2.08E+07 \\
	10 & 4d$^2$5s$^1$ & even & $^2$D & 2.5 & 45 & 4d$^2$5p$^1$ & odd & $^2$F & 3.5 & 7.66E+07 & 4.35E+07 \\
	10 & 4d$^2$5s$^1$ & even & $^2$D & 2.5 & 47 & 4d$^2$5p$^1$ & odd & $^2$D & 2.5 & 1.25E+07 & 2.84E+06 \\
	\hline
    \\
    \caption{Comparison between the A-values from the GRASP$^0$ calculation of Zr {\sc ii}, with A-values from~\protect\cite{ljung_2006}. The index numbers of the levels are from the energy order of the shifted calculation.}
    \label{zriiavaluecomparisontable}
\end{longtable}

\newpage
\begin{longtable}{ c c c c c c c c c c c c }
	\hline
	\multicolumn{5}{c}{Lower} & \multicolumn{5}{c}{upper} & \multicolumn{2}{c}{A-values (s$^{-1}$)} \\
	Level & Conf. & Parity & Term & J & Level & Conf. & Parity & Term & J & GRASP$^0$ & \cite{Reader_1997} \\
	\hline
	1 & 4d$^2$ & even & $^3$F & 2 & 15 & 4d$^1$5p$^1$ & odd & $^1$D & 2 & 7.39E+07 & 1.21E+08 \\
    1 & 4d$^2$ & even & $^3$F & 2 & 16 & 4d$^1$5p$^1$ & odd & $^3$F & 2 & 2.31E+08 & 1.23E+08 \\
    1 & 4d$^2$ & even & $^3$F & 2 & 17 & 4d$^1$5p$^1$ & odd & $^3$D & 1 & 6.86E+08 & 6.10E+08 \\
    2 & 4d$^2$ & even & $^3$F & 3 & 18 & 4d$^1$5p$^1$ & odd & $^3$F & 3 & 2.69E+08 & 2.38E+08 \\
    2 & 4d$^2$ & even & $^3$F & 3 & 19 & 4d$^1$5p$^1$ & odd & $^3$D & 2 & 6.04E+08 & 5.89E+08 \\
    2 & 4d$^2$ & even & $^3$F & 3 & 21 & 4d$^1$5p$^1$ & odd & $^3$F & 4 & 1.74E+07 & 1.60E+07 \\
    3 & 4d$^2$ & even & $^3$F & 4 & 18 & 4d$^1$5p$^1$ & odd & $^3$F & 3 & 2.69E+08 & 5.29e+06 \\
    3 & 4d$^2$ & even & $^3$F & 4 & 20 & 4d$^1$5p$^1$ & odd & $^3$D & 3 & 1.89E+06 & 6.04e+08 \\
    3 & 4d$^2$ & even & $^3$F & 4 & 21 & 4d$^1$5p$^1$ & odd & $^3$F & 4 & 1.74E+07 & 2.30e+08 \\
    4 & 4d$^2$ & even & $^1$D & 2 & 15 & 4d$^1$5p$^1$ & odd & $^1$D & 2 & 5.47E+08 & 3.35E+08 \\
    4 & 4d$^2$ & even & $^1$D & 2 & 16 & 4d$^1$5p$^1$ & odd & $^3$F & 2 & 1.26E+08 & 2.28E+08 \\
    5 & 4d$^2$ & even & $^3$P & 0 & 17 & 4d$^1$5p$^1$ & odd & $^3$D & 1 & 3.72E+07 & 4.06E+07 \\
    6 & 4d$^2$ & even & $^3$P & 1 & 17 & 4d$^1$5p$^1$ & odd & $^3$D & 1 & 1.50E+07 & 1.18E+07 \\
    6 & 4d$^2$ & even & $^3$P & 1 & 19 & 4d$^1$5p$^1$ & odd & $^3$D & 2 & 3.58E+07 & 4.98E+07 \\
    7 & 4d$^2$ & even & $^3$P & 2 & 15 & 4d$^1$5p$^1$ & odd & $^1$D & 2 & 1.87E+06 & 2.53E+06 \\
    7 & 4d$^2$ & even & $^3$P & 2 & 16 & 4d$^1$5p$^1$ & odd & $^3$F & 2 & 2.36E+06 & 1.95E+06 \\
    7 & 4d$^2$ & even & $^3$P & 2 & 20 & 4d$^1$5p$^1$ & odd & $^3$D & 3 & 3.99E+07 & 4.89E+07 \\
    8 & 4d$^2$ & even & $^1$G & 4 & 18 & 4d$^1$5p$^1$ & odd & $^3$F & 3 & 8.11E+05 & 9.66E+05 \\
    8 & 4d$^2$ & even & $^1$G & 4 & 20 & 4d$^1$5p$^1$ & odd & $^3$D & 3 & 1.36E+06 & 4.68E+06 \\
    8 & 4d$^2$ & even & $^1$G & 4 & 21 & 4d$^1$5p$^1$ & odd & $^3$F & 4 & 1.67E+05 & 2.40E+05 \\
    9 & 4d$^1$5s$^1$ & even & $^3$D & 1 & 15 & 4d$^1$5p$^1$ & odd & $^1$D & 2 & 2.90E+07 & 6.55E+07 \\
    9 & 4d$^1$5s$^1$ & even & $^3$D & 1 & 16 & 4d$^1$5p$^1$ & odd & $^3$F & 2 & 1.18E+08 & 1.73E+08 \\
    9 & 4d$^1$5s$^1$ & even & $^3$D & 1 & 17 & 4d$^1$5p$^1$ & odd & $^3$D & 1 & 2.66E+08 & 2.43E+08 \\
    9 & 4d$^1$5s$^1$ & even & $^3$D & 1 & 19 & 4d$^1$5p$^1$ & odd & $^3$D & 2 & 2.31E+08 & 1.19E+08 \\
    9 & 4d$^1$5s$^1$ & even & $^3$D & 1 & 22 & 4d$^1$5p$^1$ & odd & $^3$P & 1 & 1.34E+08 & 1.63E+08 \\
    9 & 4d$^1$5s$^1$ & even & $^3$D & 1 & 23 & 4d$^1$5p$^1$ & odd & $^3$P & 0 & 4.06E+08 & 4.27E+08 \\
    10 & 4d$^1$5s$^1$ & even & $^3$D & 2 & 18 & 4d$^1$5p$^1$ & odd & $^3$F & 3 & 1.48E+08 & 2.37E+08 \\
    10 & 4d$^1$5s$^1$ & even & $^3$D & 2 & 19 & 4d$^1$5p$^1$ & odd & $^3$D & 2 & 1.20E+08 & 1.95E+08 \\
    10 & 4d$^1$5s$^1$ & even & $^3$D & 2 & 20 & 4d$^1$5p$^1$ & odd & $^3$D & 3 & 2.50E+08 & 1.42E+08 \\
	\hline
    \\
    \caption{Comparison between the A-values from the GRASP$^0$ calculation of Zr {\sc iii}, with A-values from~\protect\cite{Reader_1997}. The index numbers of the levels are from the energy order of the shifted calculation.}
    \label{zriiiavaluecomparisontable}
\end{longtable}

\twocolumn

\section{Electron-Impact Excitation}\label{eiesection}

Using the parallel version of the Dirac Atomic R-matrix code {\sc {\sc darc}} \citep{rmatrix,Norrington_1987} electron-impact excitation calculations for Zr {\sc i} - {\sc iii} were performed using the structure models outlined in Sec.~\ref{structuresection}. A discussion of each of these three collision calculations is given below.

For Zr {\sc i}, 150 levels were maintained in the close-coupling expansion out of a a possible 726 levels included in the CI description of the target. This number was chosen to keep the computations to a manageable size, to ensure that the energies of the target states could be shifted to their experimentally known positions and also to ensure that the relevant KNe temperature range was spanned. The calculations included 78 partial waves with a total angular momentum of $2J=1-77$ with a maximum of 924 channels and a maximum Hamiltonian matrix size of 28028. A continuum basis size of 30 was adopted in the computations. The grid of incident electron energies over which the collision strengths were computed for the low partial waves ($2J=1-29$) ranged from 0.0 - 4.0~Ryds, where 0.0 - 0.6~Ryds was a fine mesh spacing of 0.00002~Ryds representing 30,000 individual energy points and 0.6 - 4.0~Ryds used a coarser mesh with a spacing of 0.0034~Ryds including 1000 energy points. For the higher partial waves ($2J=31-77$) where the resonance contributions are considerably less, a coarse mesh of 0.01~Ryds ranging from 0.0 - 4.0~Ryds. A top-up procedure \citep{Burgess_1974} was applied to account for contributions from partial waves $2J\ge 77$, important to ensure that all contributions to the slowing converging allowed lines were included. Infinite energy points for the dipoles were also computed for completeness. It is implicit that the calculation of infinite energy points and the top-up procedure is employed for the remaining two ion stages.

The comparable calculations for Zr {\sc ii} included all of the possible 241 target levels from the set of target configurations. In the electron-impact scattering model, 78 partial waves with total angular momentum $2J=0-76$ were included, with a continuum basis size of 20, a maximum of 1460 channels, resulting and a maximum Hamiltonian matrix size of 29230. A fine mesh of energies with spacing 0.00003125~Ryds was used for the low partial waves between 0.0 - 1.0~Ryds and a coarser mesh of 0.003~Ryds between 1.0 - 4.0~Ryds, with a coarse mesh of 0.004~Ryds between 0.0 - 4.0~Ryds for the higher partial waves.

Finally, for Zr {\sc iii} 167 out of the possible 265 target levels were included, with 78 partial waves with $2J=1-77$, a continuum basis size of 30, a maximum of 917 channels, and a maximum matrix size of 43701. A fine mesh of 0.00004~Ryds and a coarse grid with spacing 0.002~Ryds were adopted for the low partial waves between 0.0 - 2.0~Ryds and 2.0 - 4.0~Ryds respectively, with a coarse mesh of 0.01~Ryds between 0.0 - 4.0~Ryds for the higher partial waves.

\begin{figure*}
    \centering
    \includegraphics[width=0.9\textwidth]{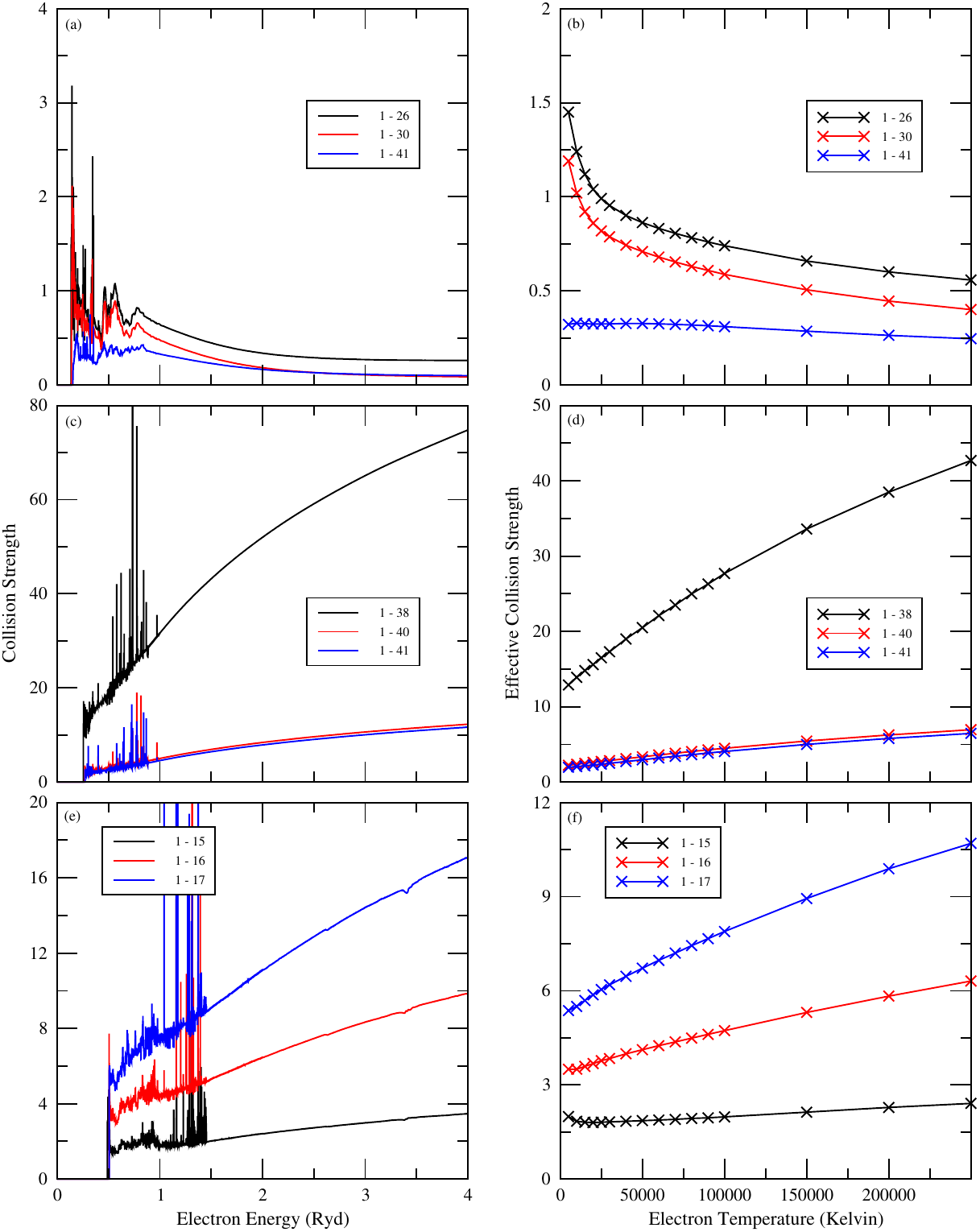}
    \caption{Collision strengths (left panels) and effective collision strengths (right panels) for three strong dipole transitions in Zr {\sc i}-{\sc iii}. Results for Zr {\sc i} are plotted in (a) and (b) for transitions 1-26, 1-30 and 1-41, Zr {\sc ii} are shown in (c) and (d) for transitions 1-38, 1-40 and 1-41, and Zr {\sc iii} are shown in (e) and (f) for transitions 1-15, 1-16 and 1-17. 
    The level indices correspond to those listed in Tables~\ref{adf04energylevelszri},~\ref{adf04energylevelszrii} and~\ref{adf04energylevelszriii} respectively.}
    \label{zrmutiplot}
\end{figure*}

Three of the lowest-lying, strongest dipole transitions from the ground state for each species have been selected as representative examples from the collision calculations. For Zr {\sc i} the transitions 1-26, 1-30 and 1-41 were chosen, these indexes are from Table~\ref{adf04energylevelszri} and are for transitions among levels 1: 4d$^2$5s$^2$ $^3$F$_2$, 26: 4d$^2$5s5p $^5$G$_2$, 30: 4d$^2$5s5p $^5$G$_3$ and 41: 4d$^2$5s5p $^5$F$_1$. The collision strengths as a function of incident electron energy in Ryds for these three transitions are shown in Fig.~\ref{zrmutiplot}a and the corresponding Maxwellian averaged effective collision strengths are shown in Fig.~\ref{zrmutiplot}b as a function of electron temperature in Kelvin. Clearly evident are the Rydberg resonances converging onto the target state thresholds in the low-energy region, followed by the smooth background cross section at higher energies. For these spin-changing M1 lines the collision strengths remain relatively constant for energies above approximately 2 Ryds. The corresponding effective collision strengths, depicted in Fig.~\ref{zrmutiplot}b replicate this behaviour, the resonance features causing an enhancement at low temperatures with the effective rates reaching a near constant value for the higher temperatures considered.

For Zr {\sc ii} the transitions chosen were 1-38, 1-40 and 1-41, these indexes are from Table~\ref{adf04energylevelszrii} and are for transitions among levels 1: 4d$^2$5s $^4$F$_{3/2}$, 38: 4d$^2$5p $^4$G$_{5/2}$, 40: 4d$^2$5p $^2$F$_{5/2}$ and 41: 4d$^2$5p $^4$F$_{3/2}$. Transition 1-38 represents a strong E1 dipole and the collision strength for this transition, shown in Fig.~\ref{zrmutiplot}c (black) exhibits the correct behaviour of a steep rise towards the high energy limit as the incident electron energy increases. The corresponding effective collision strength shown in Fig.~\ref{zrmutiplot}d mirrors this trend. Transitions 1-40 and 1-41 represent spin-changing M1 dipoles for which the collision strengths show a less pronounced rise of the background cross section at higher energies, as expected. 
Finally, for Zr {\sc iii} the collision strengths for transitions 1-15, 1-16 and 1-17 are plotted in Fig.~\ref{zrmutiplot}e, and the corresponding effective collision strengths in Fig.~\ref{zrmutiplot}f. These indexes are from Table~\ref{adf04energylevelszriii} and are for transitions between levels 1: 4d$^2$ $^3$F$_2$, 15: 4d5p $^1$D$_2$, 16: 4d5p $^3$F$_2$ and 17: 4d5p $^3$D$_1$. 

\begin{figure*}
    \centering
    \includegraphics[width=\textwidth]{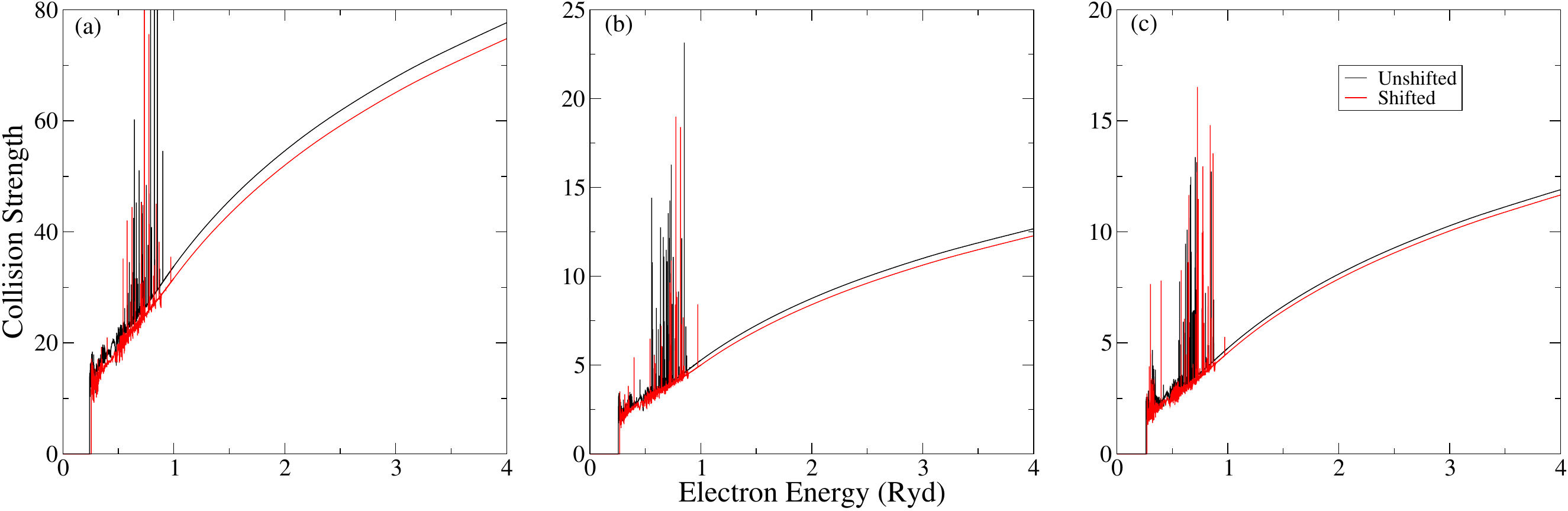}
    \caption{Collision strengths as a function of incident electron energy in Ryds for three transitions of Zr {\sc ii}. Panel (a) transition 1-38 (4d$^2$5s $^4$F$_{3/2}$- 4d$^2$5p $^4$G$_{5/2}$), panel (b) transition 1-40 (4d$^2$5s $^4$F$_{3/2}$-4d$^2$5p $^2$F$_{5/2}$) and panel (c) transition 1-41 (4d$^2$5s $^4$F$_{3/2}$- 4d$^2$5p $^4$F$_{3/2}$). The black curves represent the collision strengths computed when the {\it ab initio} energies are adopted in the collision calculations and the red curves represent the collision strengths computed with calibrated experimental energy levels.}
    \label{shiftedmutiplot}
\end{figure*}

In Sec.~\ref{structuresection} the effect on the A-values due to calibrating the energy levels to their measured experimental positions was carefully analysed. For consistency a similar investigation is carried out here to examine the effect, if any, on the collision strengths and effective collision strengths due to implementing an energy shift on the target level positions during the electron-impact excitation computations. The impact of this is shown in Fig.~\ref{shiftedmutiplot} for the same three example transitions for Zr {\sc ii} plotted in Fig.~\ref{zrmutiplot}. Transition 1-38 (4d$^2$5s $^4$F$_{3/2}$-4d$^2$5p $^4$G$_{5/2}$) is shown in Fig.~\ref{shiftedmutiplot}a, transition 1-40 (4d$^2$5s $^4$F$_{3/2}$-4d$^2$5p $^2$F$_{5/2}$) in Fig.~\ref{shiftedmutiplot}b and finally transition 1-41 (4d$^2$5s $^4$F$_{3/2}$-4d$^2$5p $^4$F$_{3/2}$) in Fig.~\ref{shiftedmutiplot}c. From these comparisons we can conclude that the effect of fine-tuning the diagonal elements of the energy Hamiltonian during the collision calculation has little effect on the final collision strengths at all incident electron energies, and hence will not significantly alter the corresponding effective collision strengths. A considerable effort was made to scrutinise a large selection of dipole and quadrupole transitions for all three charge states of Zr to ensure that this statement was corroborated. 

\section{Collisional Modelling}\label{modellingsection}

\begin{figure*}
    \centering
    \includegraphics[width=\textwidth]{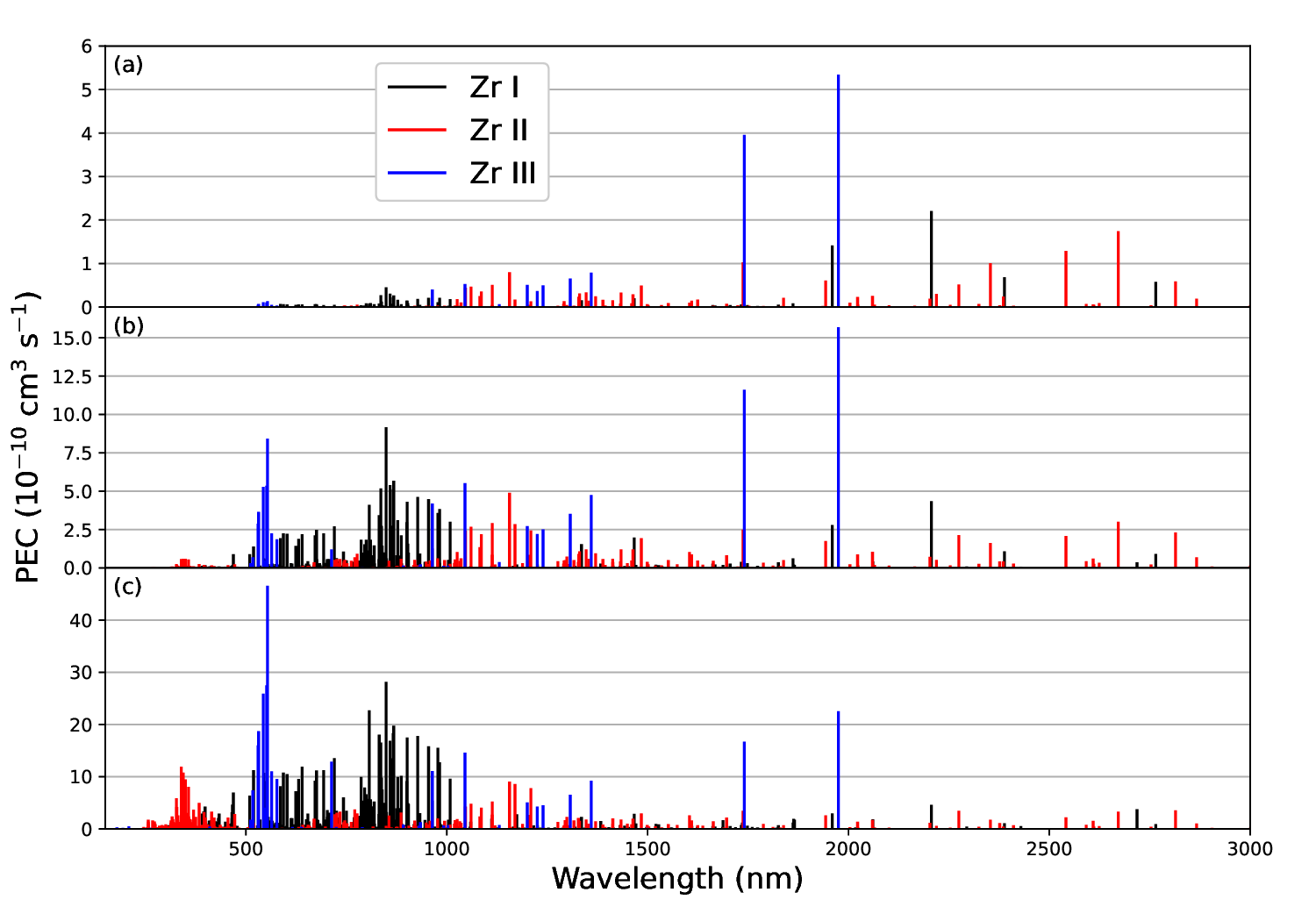}
    \caption{Theoretical spectra for Zr~I-III with an electron density of $1\times10^6$~cm$^{-3}$, where the top window (a) is for an electron energy of 0.25~eV, the middle window (b) is for 0.5~eV and the bottom window (c) is for 1~eV.}
    \label{zrspectra_tempvar}
\end{figure*}

Using the Python {\sc ColRadPy} codes \citep{colradpy}, based on the collisional radiative theory of \cite{Summers2006}, synthetic spectra can be generated for Zr using this newly calculated atomic data. In this section we concentrate on the computation of the excitation photon emissivity coefficients (PECs), often used by modellers to predict individual spectral line emission features. A PEC is a derived coefficient that is associated with a single spectral line and its excitation component (the full PEC is a combination of excitation, recombination and change exchange components) is given by 
\begin{equation}
    \mathrm{PEC}_{j \rightarrow i}^{\mathrm{excit}} = \frac{N_j^{\mathrm{excit}} A_{j \rightarrow i}}{n_e},
\end{equation} 
where $N_j^{\mathrm{excit}}$ is the weighted population of the upper level $j$ defined so that
\begin{equation}
N_{j}^{\mathrm{excit}} = \frac{N_j}{N_1},
\end{equation}
where $N_j$ is the population of the upper level and $N_1$ is the population of the ground state. As defined previously $A_{j \rightarrow i}$ is the Einstein A-coefficient for the transition from $j$ to $i$ and $n_e$ is the electron density in cm$^{-3}$. In Fig.~\ref{zrspectra_tempvar} we present example synthetic spectra computed at three electron temperatures (0.25~eV, 0.5~eV and 1~eV) and an electron density of $1\times10^6$~cm$^{-3}$ relevant for KNe modelling. In this figure the PECs (in units cm$^{3}$ s$^{-1}$) are plotted for each of the three ion stages of Zr and the wavelength region considered ranges from 0 to 3000~nm. This range is chosen to show the bulk of the lines and important lines at longer wavelengths are discussed later, only the relative heights of lines within the same ionisation stage should be considered. Below 1000~nm a forest of Zr {\sc i}, {\sc ii} and {\sc iii} lines are visible and it is in this region that most of the neutral Zr lines are found. Above 1000~nm prominent Zr {\sc ii} and {\sc iii} features are evident and appear unblended. 

\begin{figure*}
    \centering
    \includegraphics[width=\textwidth]{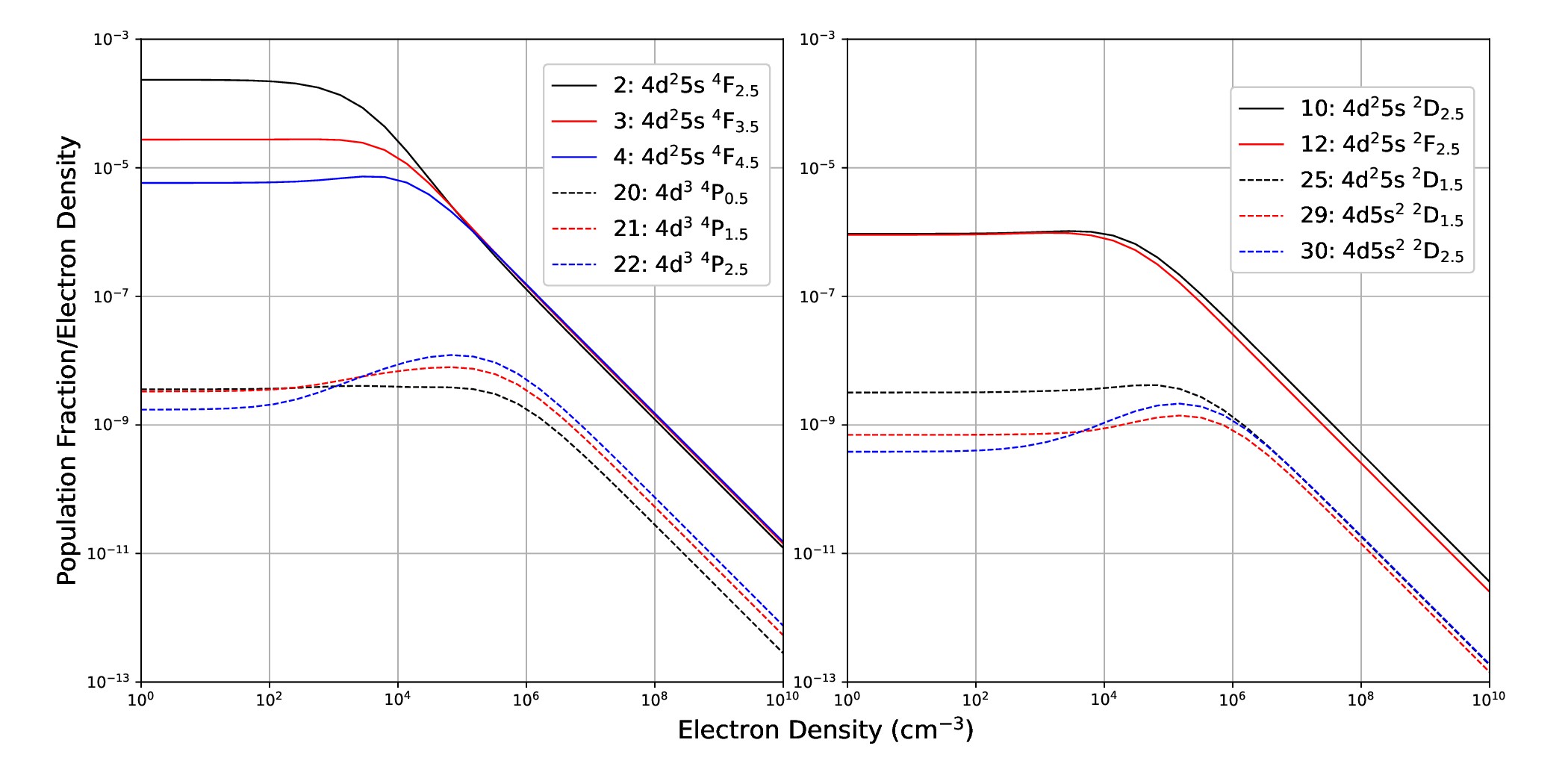}
    \caption{Electron density dependence of the population fractions of the metastable levels involved in transitions listed in Table~\ref{zrii_lines}. Calculated for an electron temperature of 0.43~eV ($\sim$5000~Kelvin).}
    \label{lte_density_plot1}
\end{figure*}

To further analyse these findings, we refer to a paper by \cite{Gillanders_2024} in which several Zr {\sc ii} lines were highlighted in the 1.0 -- 1.2~micron range. In that publication Table C1 listed several Zr {\sc ii} lines of interest, the data for which came from the Kurucz atomic database \cite{kurucz} and lines were selected based on their calculated luminosity. We have extracted this list and present them in Table~\ref{zrii_lines} highlighting the relevant wavelengths, the lower and upper index labels and energies (in eV for consistency) computed in the current calculations. The associated A-values and the relative intensities of the lines that were calculated from the new data presented in this work are compared to the values in \cite{Gillanders_2024}, where an electron temperature of 0.43~eV ($\sim$5000~Kelvin) and density of $1\times10^6$~cm$^{-3}$ were used. Under these conditions there is agreement for the lines shown suggesting that the assumptions made in \cite{Gillanders_2024} were reasonable.
\begin{table*}
	\centering
	\begin{tabular}{ c c c c c c c c c }
		\hline
        Wavelength & \multicolumn{2}{c}{Lower} & \multicolumn{2}{c}{Upper} & \multicolumn{2}{c}{A-value (s$^{-1}$)} & \multicolumn{2}{c}{Relative Intensity} \\
        (nm) & Index & eV & Index & eV & This work & G24 & This work & G24 \\
		\hline
        1156.6 & 4 & 0.164 & 22 & 1.236 & 0.074 & 0.079 & 1.00 & 1.00 \\
        1113.7 & 3 & 0.095 & 21 & 1.208 & 0.065 & 0.066 & 0.60 & 0.62 \\
        1060.7 & 2 & 0.039 & 21 & 1.208 & 0.060 & 0.059 & 0.55 & 0.58 \\
        1046.8 & 1 & 0.000 & 20 & 1.184 & 0.107 & 0.105 & 0.50 & 0.55 \\
        1170.2 & 12 & 0.713 & 29 & 1.773 & 0.223 & 0.110 & 0.49 & 0.26 \\
        1086.3 & 3 & 0.095 & 22 & 1.236 & 0.033 & 0.035 & 0.45 & 0.47 \\
        1082.4 & 2 & 0.039 & 20 & 1.184 & 0.060 & 0.060 & 0.28 & 0.31 \\
        1026.4 & 1 & 0.000 & 21 & 1.208 & 0.023 & 0.023 & 0.21 & 0.23 \\
        1113.5 & 12& 0.713 & 30 & 1.827 & 0.046 & 0.038 & 0.14 & 0.13 \\
        1035.8 & 2 & 0.039 & 22 & 1.236 & 0.009 & 0.010 & 0.13 & 0.14 \\
        1021.1 & 10 & 0.559 & 29 & 1.773 & 0.043 & 0.056 & 0.10 & 0.15 \\
        1120.7 & 10 & 0.559 & 25 & 1.665 & 0.011 & 0.034 & 0.04 & 0.11 \\
        \hline
	\end{tabular}
    \caption{Zr~II lines highlighted in \protect\cite{Gillanders_2024} (G24), where the index numbers represent the indexes of the levels from the Zr~II calculation in this work. A comparison is made between this work and the data from \protect\cite{kurucz} at an electron temperature of of 0.43~eV ($\sim$5000~Kelvin) and density of $1\times10^6$~cm$^{-3}$, showing reasonable agreement for these lines.}
    \label{zrii_lines}
\end{table*}
The assumption of LTE can be probed by investigating the electron density dependence of the populations in these levels. This can be done using the collisional radiative theory in {\sc ColRadPy}, allowing for a determination of when these levels leave LTE and enter a NLTE regime. In Fig.~\ref{lte_density_plot1} the ratio of the population fractions and the electron density are plotted as a function of electron density (cm$^{-3}$) for the levels indexed in Table~\ref{zrii_lines}. Clearly we can see the evolution of the population fractions as the electron density increases. When the gradient of the curve in the logarithmic plots becomes -1 the level it corresponds to is determined to be in LTE. From Fig.~\ref{lte_density_plot1} it appears all the metastable levels considered are in NLTE for electron densities below approximately $1\times10^6$~cm$^{-3}$ and enter LTE for electron densities greater than this. Therefore it is likely that a fully NLTE description will be necessary at later times in the KNe evolution, as the density rapidly decreases due to the ejecta expansion.



\begin{figure*}
    \centering
    \includegraphics[width=\textwidth]{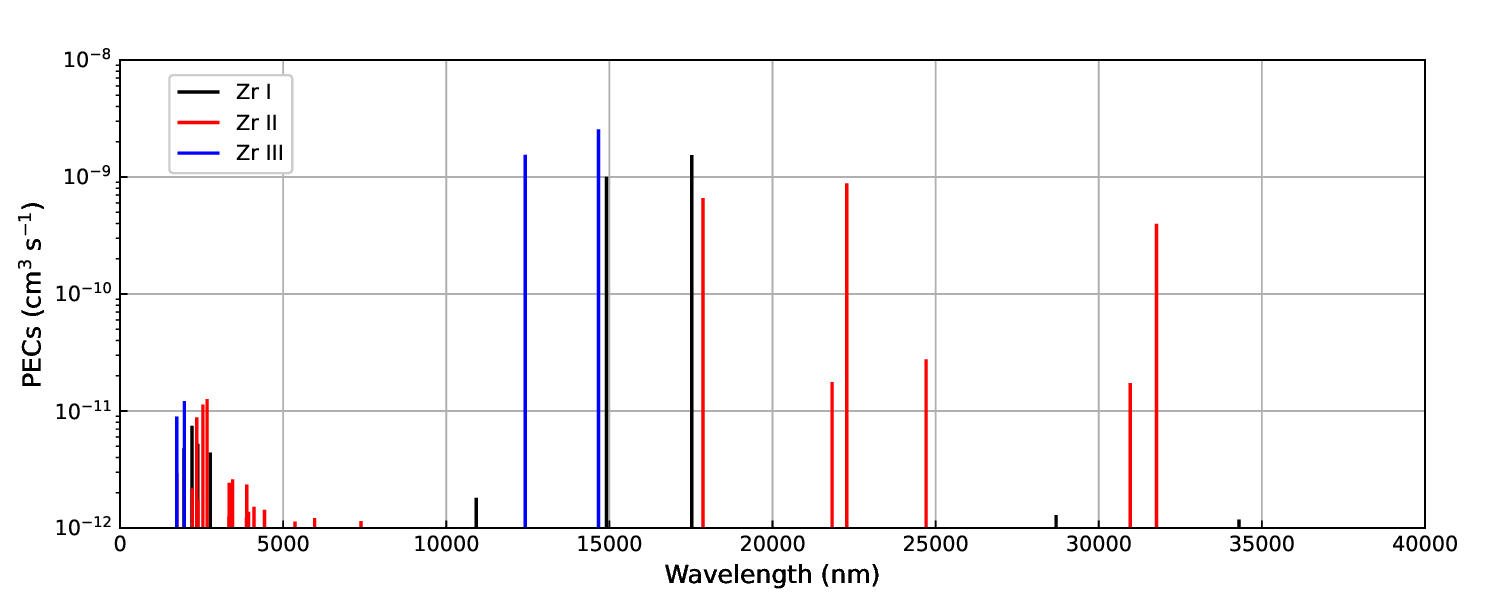}
    \caption{Theoretical spectra for Zr~I-III with an electron temperature of 0.1~eV (1160~Kelvin) and electron density of $1\times10^6$~cm$^{-3}$. This temperature is more relevant to later stages in the merger evolution, where a few lines have a relative intensity significantly greater than all the rest.}
    \label{zrspectra_lowtemp}
\end{figure*}

At later stages in the evolution of the KNe the temperature decreases to such an extent that only the lowest lying target levels can be excited. To study this further we plot in Fig.~\ref{zrspectra_lowtemp} the PECs (in units cm$^{3}$s$^{-1}$) computed at a low temperature of 0.1~eV (1160~Kelvin) for an extended wavelength region from 0 to 40000~nm. Clearly evident are lines between these low-lying levels which have a much greater relative intensity than the lines previously presented in Fig.~\ref{zrspectra_tempvar} between 0 and 3000~nm. Prominent features of Zr {\sc i} and particularly Zr {\sc ii} and Zr {\sc iii} are evident at much higher wavelengths. The relevant lines are listed in Table~\ref{zr_lowtemplines} including the wavelength, ion species, lower and upper index from the present computations and the magnitude of the PEC. All of these lines represent transitions among the lowest 4 levels of the relevant Zr charge species.

\begin{table}
	\centering
	\begin{tabular}{ c c c c c }
		\hline
        Wavelength & Ion & Lower & Upper & PECs \\
        (nm) & & Index & Index & (cm$^3$s$^{-1}$) \\
		\hline
        12424.56 & Zr~III & 2 & 3 & 1.54E-09 \\
        14671.59 & Zr~III & 1 & 2 & 2.56E-09 \\
        14915.82 & Zr~I & 2 & 3 & 1.00E-09 \\
        17531.11 & Zr~I & 1 & 2 & 1.53E-09 \\
        17874.29 & Zr~II & 3 & 4 & 6.58E-10 \\
        22283.09 & Zr~II & 2 & 3 & 8.80E-10 \\
        31779.15 & Zr~II & 1 & 2 & 3.98E-10 \\
        \hline
	\end{tabular}
    \caption{The seven lines with the largest PECs presented in Fig.~\ref{zrspectra_lowtemp}, computed at an electron temperature of 0.1~eV and an electron density of $1\times10^6$~cm$^{-3}$.}
    \label{zr_lowtemplines}
\end{table}

The new atomic data produced in this work can be used to predict luminosity values for prominent emission features found in the synthetic spectra. These luminosities are calculated in terms of the previously discussed PECs and are defined as,
\begin{equation}
    L_{j \to i} = \frac{hc}{\lambda_{j \to i}}   \frac{n_e\text{PEC}_{j\to i } }{\sum_i N_i} \frac{M_{\text{ion}}}{m_{\text{ion}}},
    \label{eq:lumo}
\end{equation}
in units of energy per time, where $hc/\lambda$ is the photon energy, $M_{\text{ion}}$ is the mass of the ion in the ejecta and $m_{\text{ion}}$ is the mass of a single ion particle. The ratio of $M_{\text{ion}}$ and $m_{\text{ion}}$ encodes the number of ions in the ejecta. We present in Table~\ref{tab:lumozr} luminosity predictions for the ten most intense lines of Zr {\sc i}, {\sc ii} and {\sc iii} computed at an electron temperature of 0.25~eV, an electron density of $1\times10^{6}$~cm$^{-3}$ and an ejecta mass of $1\times10^{-3} M_\odot$. 
It should be noted that the PEC values for these transitions have changed somewhat compared to Table~\ref{zr_lowtemplines}, as the computations in Table~\ref{tab:lumozr} were performed at a higher electron temperature of 0.25~eV. We also note that the luminosity values for each species are of the same order of magnitude for all transitions, $\approx 10^{36}$~erg s$^{-1}$. 
These luminosity predictions for Zr can be useful when making ion-mass estimates. Additionally, referring back to Table~\ref{zrii_lines} the first four highest intensity lines between 1.0 -- 1.2~micron in \cite{Gillanders_2024} are also included in the 10 strongest lines for Zr~{\sc ii} predicted in Table~\ref{tab:lumozr}. 

\begin{table*}
	\begin{tabular}{crcrrrrrrlcc}
		\hline
		& $\lambda$ &Index         & \multicolumn{3}{c}{Lower Level} & \multicolumn{3}{c}{Lower Level} & {$A_{{j\to i}}$} & PEC              & Luminosity      \\  
		& (nm)      &(${i}$-${j}$) & Conf. & J & E (cm$^{-1}$)    & Conf. & J & E (cm$^{-1}$)    & (s$^{-1}$)       & (cm$^3$s$^{-1}$) & (erg s$^{-1}$ ) \\
		\hline	
		\multirow{10}{*}{\rotatebox[origin=c]{90}{Zr~I}} &2206.89&     2 -  9&   4d$^2$5s$^2$& 3.0& 570.414& 4d$^2$5s$^2$& 2.0& 5101.676& 1.56E-02&	    2.25E-10& 2.64E+36 \\
 		&14915.82&    2 -  3&   4d$^2$5s$^2$& 3.0& 570.414& 4d$^2$5s$^2$& 4.0& 1240.843& 6.09E-03&	    1.42E-09& 2.47E+36 \\
 		&17531.11&    1 -  2&   4d$^2$5s$^2$& 2.0&   0.000& 4d$^2$5s$^2$& 3.0&  570.414& 4.74E-03&	    1.34E-09& 1.99E+36 \\
  		&1960.14&     1 -  9&   4d$^2$5s$^2$& 2.0&   0.000& 4d$^2$5s$^2$& 2.0& 5101.676& 1.00E-02&	    1.44E-10& 1.91E+36 \\
   		&849.48&      2 - 19&   4d$^2$5s$^2$& 3.0& 570.414& 4d$^3$5s$^1$& 4.0&12342.374& 9.32E-01&	    4.48E-11& 1.37E+36 \\
   		&859.05&      1 - 17&   4d$^2$5s$^2$& 2.0&   0.000& 4d$^3$5s$^1$& 2.0&11640.713& 1.28E+00&	    3.49E-11& 1.05E+36 \\
   		&836.38&      1 - 18&   4d$^2$5s$^2$& 2.0&   0.000& 4d$^3$5s$^1$& 3.0&11956.329& 9.16E-01&	    3.05E-11& 9.48E+35 \\
   		&868.07&      3 - 21&   4d$^2$5s$^2$& 4.0&1240.843& 4d$^3$5s$^1$& 4.0&12760.659& 1.13E+00&	    3.11E-11& 9.30E+35 \\
   		&867.16&      3 - 22&   4d$^2$5s$^2$& 4.0&1240.843& 4d$^3$5s$^1$& 5.0&12772.774& 7.29E-01&	    2.67E-11& 7.98E+35 \\
  		&2388.86&     1 -  4&   4d$^2$5s$^2$& 2.0&   0.000& 4d$^2$5s$^2$& 2.0& 4186.105& 2.28E-03&	    6.82E-11& 7.41E+35 \\
		\hline
		\multirow{10}{*}{\rotatebox[origin=c]{90}{Zr~II}} &1156.64&	4 - 22&   4d$^2$5s$^1$& 4.5& 1322.905&     4d$^3$& 2.5&9968.646&  7.36E-02&   7.97E-11& 1.79E+36 \\
		&2672.33&	3 - 10&   4d$^2$5s$^1$& 3.5&  763.442&  4d$^2$5s$^1$& 2.5&4505.495&  7.88E-03&   1.74E-10& 1.69E+36 \\
		&1738.25&	1 - 12&   4d$^2$5s$^1$& 1.5&    0.000&  4d$^2$5s$^1$& 2.5&5752.923&  8.56E-03&   1.02E-10& 1.53E+36 \\
		&2542.18&	2 -  9&   4d$^2$5s$^1$& 2.5&  314.672&  4d$^2$5s$^1$& 1.5&4248.304&  7.70E-03&   1.29E-10& 1.32E+36 \\
		&17874.29&	3 -  4&   4d$^2$5s$^1$& 3.5&  763.442&  4d$^2$5s$^1$& 4.5&1322.905&  4.71E-03&   8.56E-10& 1.24E+36 \\
		&1113.67&	3 - 21&   4d$^2$5s$^1$& 3.5&  763.442&     4d$^3$& 1.5&9742.796&  6.48E-02&   5.06E-11& 1.18E+36 \\
		&1060.66&	2 - 21&   4d$^2$5s$^1$& 2.5&  314.672&     4d$^3$& 1.5&9742.796&  5.95E-02&   4.64E-11& 1.14E+36 \\
		&2353.88&	1 -  9&   4d$^2$5s$^1$& 1.5&    0.000&  4d$^2$5s$^1$& 1.5&4248.304&  6.01E-03&   1.01E-10& 1.11E+36 \\
		&1046.78&	1 - 20&   4d$^2$5s$^1$& 1.5&    0.000&     4d$^3$& 0.5&9553.104&  1.07E-01&   4.43E-11& 1.10E+36 \\
		&22283.09&	2 -  3&   4d$^2$5s$^1$& 2.5&  314.672&  4d$^2$5s$^1$& 3.5& 763.442&  3.90E-03&   7.61E-10& 8.86E+35 \\
		\hline
		\multirow{10}{*}{\rotatebox[origin=c]{90}{Zr~III}} &1975.58&	2 -  4&  4d$^2$& 3.0&  681.589&  4d$^2$& 2.0&  5743.387& 3.08E-02&   5.33E-10& 7.01E+36 \\
		&12424.56&	2 -  3&  4d$^2$& 3.0&  681.589&  4d$^2$& 4.0&  1486.447& 1.05E-02&   2.96E-09& 6.18E+36 \\
		&1741.13&	1 -  4&  4d$^2$& 2.0&    0.000&  4d$^2$& 2.0&  5743.387& 2.28E-02&   3.95E-10& 5.89E+36 \\
		&14671.59&	1 -  2&  4d$^2$& 2.0&    0.000&  4d$^2$& 3.0&   681.589& 8.08E-03&   2.76E-09& 4.88E+36 \\
		&1359.89&	3 -  7&  4d$^2$& 4.0& 1486.447&  4d$^2$& 2.0&  8839.967& 2.24E-02&   7.87E-11& 1.50E+36 \\
		&1045.57&	3 -  8&  4d$^2$& 4.0& 1486.447&  4d$^2$& 4.0& 11050.568& 2.46E-02&   5.28E-11& 1.31E+36 \\
		&1307.95&	2 -  6&  4d$^2$& 3.0&  681.589&  4d$^2$& 1.0&  8327.117& 2.35E-02&   6.54E-11& 1.30E+36 \\
		&1200.90&	1 -  6&  4d$^2$& 2.0&    0.000&  4d$^2$& 1.0&  8327.117& 1.82E-02&   5.07E-11& 1.09E+36 \\
		&964.42&	2 -  8&  4d$^2$& 3.0&  681.589&  4d$^2$& 4.0& 11050.568& 1.87E-02&   4.01E-11& 1.08E+36 \\
		&1240.14&	1 -  5&  4d$^2$& 2.0&    0.000&  4d$^2$& 0.0&  8063.628& 4.65E-02&   4.96E-11& 1.04E+36 \\
		\hline
	\end{tabular}
	\caption{Luminosities for the ten strongest lines from each Zr ion considered here at $T_e = 0.25$ eV, $n_e = 1\times10^6$ cm$^{-3}$ and a mass of $1\times10^{-3} M_\odot $.}
	\label{tab:lumozr}
\end{table*} 

\section{Radiative Transport}\label{sec:tardis}

1D spectral synthesis codes such as {\sc tardis} \citep{tardis, kerzendorf_software} have previously been used to aid analysis of the observed spectra of AT2017gfo and to suggest identifications of spectral features, with \cite{Vieira_2023} and \cite{Gillanders_2024} both suggesting the presence of Zr. High quality, accurate atomic data is paramount for successful radiative transfer modelling and production of synthetic spectra. In this section, we discuss the impact our new atomic calculations have on early-time spectrum synthesis modelling by comparing with spectra previously produced using earlier datasets. We compare with the work of \cite{Gillanders_2022} - hereafter referred to as G22 - following methods described by \cite{mulholland2024}. Adopting these methods, and the parameters outlined in Table 9 of that work, we produce early time spectra resembling that of AT2017gfo. We focus on early time spectra as the kilonova is still well represented by a blackbody here, an assumption important for this model. We use two atomic datasets to compare to observations, one drawn from the same sources as G22, and one where we substitute the data for Zr {\sc i} - {\sc iii} with that presented in this work. This facilitates a direct differential test of the impact of our new atomic dataset.

\begin{figure*}
    \centering
    \includegraphics[width=\textwidth]{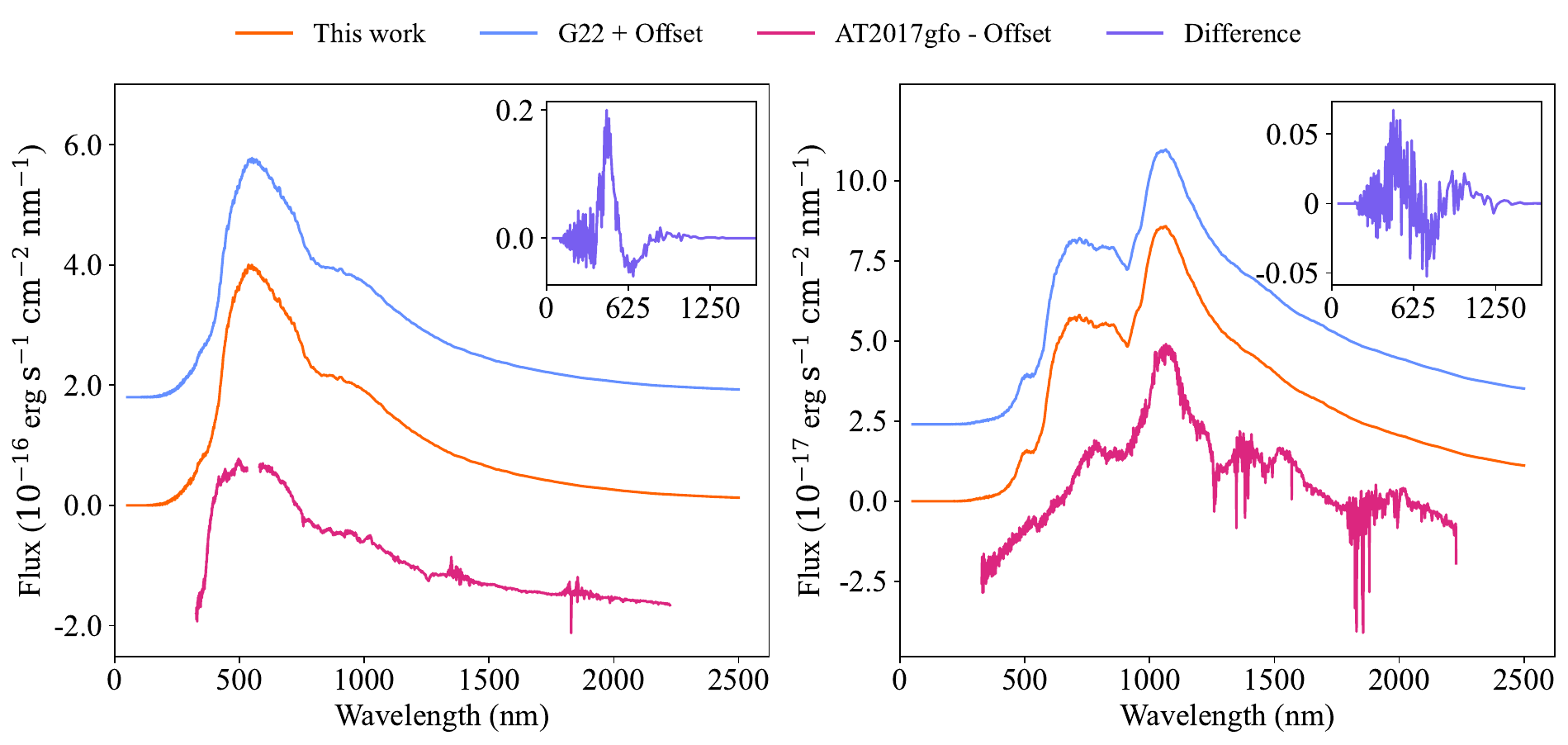}
    \caption{Line plot showing a comparison between a realistic elemental composition of AT2017gfo at 1.4 days (left) and 4.4 days (right) post-merger. The spectrum produced from the same atomic data sources as used by G22 is shown in blue, with the data set including the new calculation presented in this work for Zr shown in orange, alongside the observed spectrum \citep{pian2017, smartt2017} in pink. These observations and synthetic spectra from the G22 dataset have been arbitrarily offset from the spectrum featuring the new calculation - by $\mp$ 1.8 E-16 erg s$^{-1}$ cm$^{-2}$ nm$^{-1}$ at 1.4 days and $\mp$ 2.4 E-17 erg s$^{-1}$ cm$^{-2}$ nm$^{-1}$ at 4.4 days - for visual clarity. We note there is a visual discrepancy between the simulated spectra and the observations: this is due to using a more recent version of \textsc{tardis}, with an improved relativistic treatment \citep{vogl_2019}. Both datasets are treated with this same updated relativity to allow for comparison. An inset plot is included showing the difference between the two synthetic spectra in purple.}
    \label{fig:diff}
\end{figure*}

\begin{figure*}
    \centering
    \includegraphics[width=\textwidth]{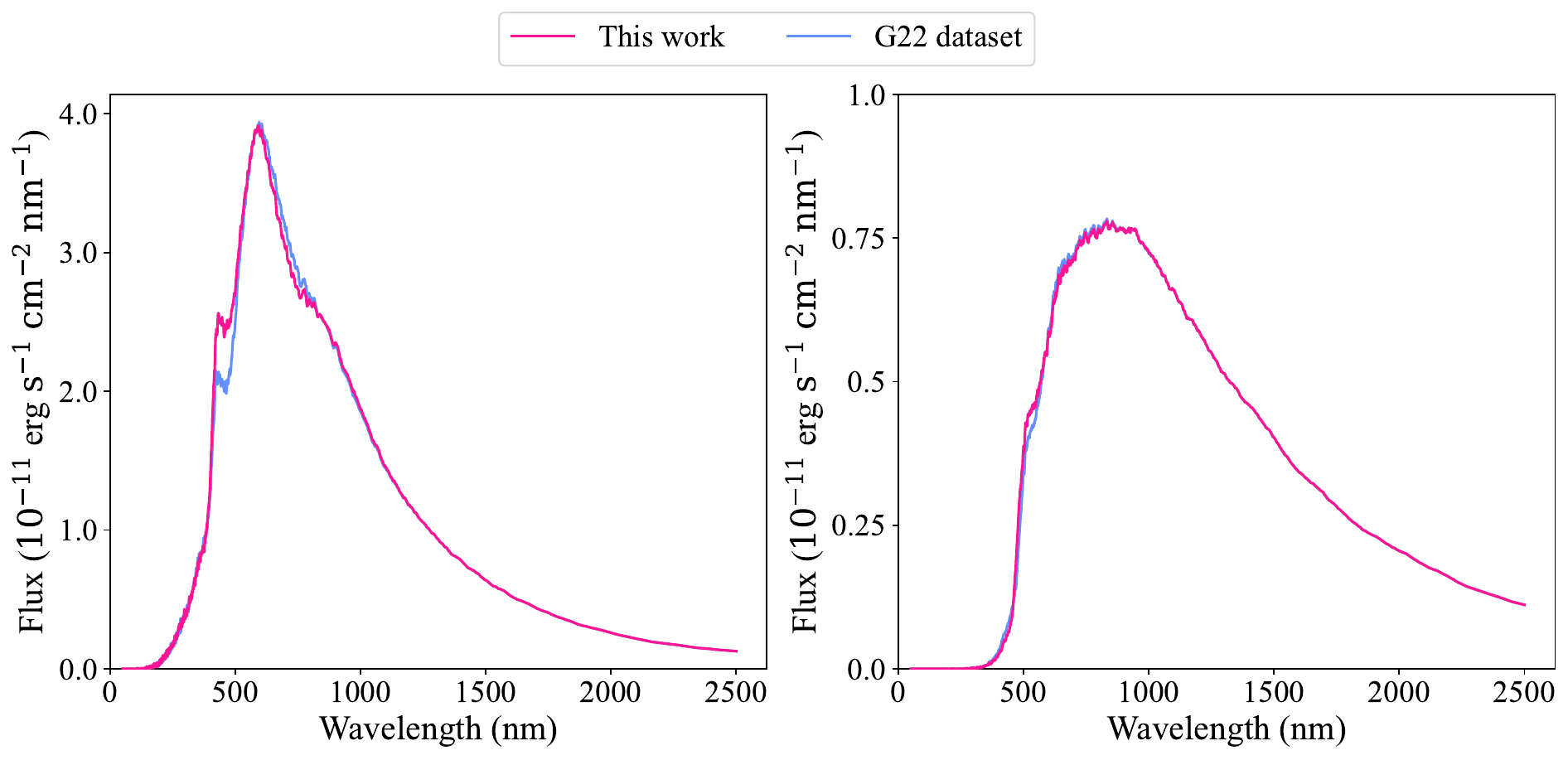}
    \caption{Comparison between pure Zr models of AT2017gfo at 1.4 days (left) and 4.4 days (right) showing differences between data used by G22 (blue) and this calculation (pink). The spectra shown here are dominated by Zr \textsc{ii}.}
    \label{fig:zr_diff}
\end{figure*}

In Fig.~\ref{fig:diff} we present a line plot of the flux (in units 10$^{-16}$ erg s$^{-1}$ cm$^{-2}$ nm$^{-1}$) as a function of wavelength (in nm), at 1.4 days (left panel) and 4.4 days (right panel) post merger. Fig.~\ref{fig:diff} shows that the addition of the new data has only a small impact on the overall spectral shape, similar to previous findings for new calculations for Sr and Y \citep{mulholland2024}. This difference is more pronounced at the earlier 1.4 day epoch, clearly seen in the inset difference plot, and is caused by changes to Zr {\sc ii} as the dominant ion stage at this time. By the 4.4 day epoch illustrated on the right in Fig.~\ref{fig:diff}, any difference between the spectra seen in the two datasets is overshadowed by the Monte Carlo noise inherent in such simulations. 

To investigate further, we construct a model with a less realistic composition, a pure Zr KNe, to isolate the effect the new data from these new calculations have on the spectra. In Fig.~\ref{fig:zr_diff}, we again see the strongest difference between the dataset used by G22 and this new simulation in the region around 500 nm at the 1.4 day observational epoch, with this difference becoming much less pronounced by the 4.4 day time frame. In summary, whilst we find only small changes in the synthetic spectra by substituting the new atomic datasets into the radiative transfer models of AT2017gfo, future further refinements to the atomic data available for such models may have a cumulative, visible effect on synthetic spectra and models, particularly with future efforts to push more towards modelling in full NLTE.

\section{Conclusions}\label{conclusionsection}

Recent interest in NLTE modelling of the KNe from NSM events have motivated this study on the structure and electron-impact excitation of the first three ion stages of Zr. Atomic structure calculations using {\sc grasp0} produced the required energy levels and transition rates, and further collision calculations using {\sc darc} provided the excitation and de-excitation rates for Zr {\sc i}, Zr {\sc ii} and Zr {\sc iii}. All these atomic data are vital for the accurate modelling of these events. 

The functionality in both the {\sc grasp0} and {\sc darc} codes to calibrate the target state energies to their exact experimental positions was adopted, thereby producing spectroscopically accurate wavelengths and therefore more accurate A-values and collision strengths. A detailed analysis of the difference between adopting shifted or unshifted energies was carried out with regard to the computation of A-values for transitions among the target levels and the corresponding electron-impact excitation collision strengths. For all species it was revealed that the most significant impact was in the resulting A-values, whereas the changes in the collision strengths was relatively small. We conclude that, if possible, target energies should be calibrated to their spectroscopic positions to maximize the accuracy of the results and aid in the identification of observed spectral features.  

The atomic data produced for Zr {\sc i}-{\sc iii} was further incorporated into a collisional radiative model using the {\sc ColRadPy} code to compute populations for the ground and metastable levels, derive PECs across a wide wavelength region and finally use these PECs to compute luminosity values for prominent line candidates in the synthetic spectra. All of this modelling was performed for electron temperatures and densities of relevance to KNe events. The transitions listed in Table~\ref{zrii_lines} were found to be in LTE for electron densities greater than $1\times10^6$~cm$^{-3}$ but for densities lower than this value NLTE modelling would be required. 

Finally the new atomic datasets were incorporated into a 1D radiative transfer simulation using the {\sc tardis} suite of computer packages. The addition of the new data had only a small impact on the overall spectral shape during photospheric phases, with the greatest differences occurring at the early observational epoch of 1.4 days post merger. By 4.4 days any differences could be attributed to Monte Carlo noise. These small differences for three lone Zr species could, however, produce a cumulative effect as more and more accurate atomic data calculations are produced in the future. This is particularly important as current efforts move more towards modelling in full NLTE.     

\section*{Acknowledgements}

Funded/Co-funded by the European Union (ERC, HEAVYMETAL, 101071865). Views and opinions expressed are however those of the author(s) only and do not necessarily reflect those of the European Union or the European Research Council. Neither the European Union nor the granting authority can be held responsible for them. This research made use of \textsc{tardis}, a community-developed software package for spectral synthesis in supernovae \citep{tardis, vogl_2019,kerzendorf_software}. The development of \textsc{tardis} received support from the Google Summer of Code initiative and from ESA’s Summer of Code in Space program. \textsc{tardis} makes extensive use of Astropy and PyNE. 

\section*{Data Availability}

The data presented in this work will be made available in the \cite{openadas} database or included with the current iteration of the R-matrix codes on the {\sc darc} website \citep{rmatrix}, in the form of adf04 files compatible with the ADAS suite of codes or the open-source ColRadPy~\citep{colradpy}.

\bibliographystyle{mnras}
\bibliography{main}

\begin{thebibliography}{}
\makeatletter
\relax
\def\mn@urlcharsother{\let\do\@makeother \do\$\do\&\do\#\do\^\do\_\do\%\do\~}
\def\mn@doi{\begingroup\mn@urlcharsother \@ifnextchar [ {\mn@doi@}
  {\mn@doi@[]}}
\def\mn@doi@[#1]#2{\def\@tempa{#1}\ifx\@tempa\@empty \href
  {http://dx.doi.org/#2} {doi:#2}\else \href {http://dx.doi.org/#2} {#1}\fi
  \endgroup}
\def\mn@eprint#1#2{\mn@eprint@#1:#2::\@nil}
\def\mn@eprint@arXiv#1{\href {http://arxiv.org/abs/#1} {{\tt arXiv:#1}}}
\def\mn@eprint@dblp#1{\href {http://dblp.uni-trier.de/rec/bibtex/#1.xml}
  {dblp:#1}}
\def\mn@eprint@#1:#2:#3:#4\@nil{\def\@tempa {#1}\def\@tempb {#2}\def\@tempc
  {#3}\ifx \@tempc \@empty \let \@tempc \@tempb \let \@tempb \@tempa \fi \ifx
  \@tempb \@empty \def\@tempb {arXiv}\fi \@ifundefined
  {mn@eprint@\@tempb}{\@tempb:\@tempc}{\expandafter \expandafter \csname
  mn@eprint@\@tempb\endcsname \expandafter{\@tempc}}}

\bibitem[\protect\citeauthoryear{Ballance}{Ballance}{2024}]{rmatrix}
Ballance C.~P.,  2024, R-matrix codes, \url{http://connorb.freeshell.org}

\bibitem[\protect\citeauthoryear{Bi\'emont, Grevesse, Hannaford  \&
  Lowe}{Bi\'emont et~al.}{1981}]{Biemont1981}
Bi\'emont E.,  Grevesse N.,  Hannaford P.,   Lowe R.~M.,  1981, \mn@doi
  [Astrophysical Journal] {https://doi.org/10.1086/159213}, 248, 867

\bibitem[\protect\citeauthoryear{Burgess}{Burgess}{1974}]{Burgess_1974}
Burgess A.,  1974, \mn@doi [Journal of Physics B: Atomic and Molecular Physics]
  {10.1088/0022-3700/7/12/003}, 7, L364

\bibitem[\protect\citeauthoryear{Burke}{Burke}{2011}]{burke2011}
Burke P.~G.,  2011, R-Matrix Theory of Atomic Collisions: Application to
  Atomic, Molecular and Optical Processes.
 Springer Series on Atomic, Optical, and Plasma Physics Vol. 61,
  Springer-Verlag Berlin Heidelberg, \url
  {https://www.springer.com/gp/book/9783642159305}

\bibitem[\protect\citeauthoryear{{Gillanders}, {Smartt}, {Sim}, {Bauswein}  \&
  {Goriely}}{{Gillanders} et~al.}{2022}]{Gillanders_2022}
{Gillanders} J.~H.,  {Smartt} S.~J.,  {Sim} S.~A.,  {Bauswein} A.,   {Goriely}
  S.,  2022, \mn@doi [\mnras] {10.1093/mnras/stac1258}, \href
  {https://ui.adsabs.harvard.edu/abs/2022MNRAS.515..631G} {515, 631}

\bibitem[\protect\citeauthoryear{Gillanders, Sim, Smartt, Goriely  \&
  Bauswein}{Gillanders et~al.}{2024}]{Gillanders_2024}
Gillanders J.~H.,  Sim S.~A.,  Smartt S.~J.,  Goriely S.,   Bauswein A.,  2024,
  \mn@doi [Monthly Notices of the Royal Astronomical Society]
  {10.1093/mnras/stad3688}, 529, 2918

\bibitem[\protect\citeauthoryear{Johnson, Loch  \& Ennis}{Johnson
  et~al.}{2019}]{colradpy}
Johnson C.,  Loch S.,   Ennis D.,  2019, \mn@doi [Nuclear Materials and Energy]
  {https://doi.org/10.1016/j.nme.2019.01.013}, 20, 100579

\bibitem[\protect\citeauthoryear{Kajino, Aoki, Balantekin, Diehl, Famiano  \&
  Mathews}{Kajino et~al.}{2019}]{kajino2019}
Kajino T.,  Aoki W.,  Balantekin A.,  Diehl R.,  Famiano M.,   Mathews G.,
  2019, \mn@doi [Progress in Particle and Nuclear Physics]
  {https://doi.org/10.1016/j.ppnp.2019.02.008}, 107, 109

\bibitem[\protect\citeauthoryear{Kerzendorf \& Sim}{Kerzendorf \&
  Sim}{2014}]{tardis}
Kerzendorf W.~E.,  Sim S.~A.,  2014, \mn@doi [Monthly Notices of the Royal
  Astronomical Society] {10.1093/mnras/stu055}, 440, 387

\bibitem[\protect\citeauthoryear{Kerzendorf et~al.,}{Kerzendorf
  et~al.}{2023}]{kerzendorf_software}
Kerzendorf W.,  et~al., 2023, tardis-sn/tardis: TARDIS v2023.08.13,
  \mn@doi{10.5281/zenodo.8244935}, \url
  {https://doi.org/10.5281/zenodo.8244935}

\bibitem[\protect\citeauthoryear{Kramida, Ralchenko, Reader  \& {NIST ASD
  Team}}{Kramida et~al.}{2024}]{nist}
Kramida A.,  Ralchenko Y.,  Reader J.,   {NIST ASD Team} 2024, NIST Atomic
  Spectra Database (version 5.11), \url{https://physics.nist.gov/asd}

\bibitem[\protect\citeauthoryear{Kurucz}{Kurucz}{2024}]{kurucz}
Kurucz R.~L.,  2024, Kurucz Atomic Database (version 5.11),
  \url{http://kurucz.harvard.edu/atoms.html}

\bibitem[\protect\citeauthoryear{Lawler, Schmidt  \& Hartog}{Lawler
  et~al.}{2022}]{Lawler2022}
Lawler J.~E.,  Schmidt J.~R.,   Hartog E. A.~D.,  2022, \mn@doi [Journal of
  Quantitative Spectroscopy and Radiative Transfer]
  {https://doi.org/10.1016/j.jqsrt.2022.108283}, 289, 108283

\bibitem[\protect\citeauthoryear{Ljung, Nilsson, Asplund  \& Johansson}{Ljung
  et~al.}{2006}]{ljung_2006}
Ljung G.,  Nilsson H.,  Asplund M.,   Johansson S.,  2006, \mn@doi [A&A]
  {10.1051/0004-6361:20065212}, 456, 1181

\bibitem[\protect\citeauthoryear{Moore}{Moore}{1971}]{Moore_1971}
Moore C.~E.,  1971, in , Nat. Stand. Ref. Data Ser., NSRDS-NBS 35, Vol. II
  (Reprint of NBS Circ. 467, Vol. II, 1952).
Nat. Bur. Stand., U.S., \mn@doi{10.6028/NBS.NSRDS.35v2}

\bibitem[\protect\citeauthoryear{Mulholland, McElroy, McNeill, Sim, Ballance
  \& Ramsbottom}{Mulholland et~al.}{2024}]{mulholland2024}
Mulholland L.~P.,  McElroy N.~E.,  McNeill F.~L.,  Sim S.~A.,  Ballance C.~P.,
   Ramsbottom C.~A.,  2024, \mn@doi [Monthly Notices of the Royal Astronomical
  Society] {10.1093/mnras/stae1615}, 532, 2289

\bibitem[\protect\citeauthoryear{Norrington \& Grant}{Norrington \&
  Grant}{1987}]{Norrington_1987}
Norrington P.~H.,  Grant I.~P.,  1987, \mn@doi [Journal of Physics B: Atomic
  and Molecular Physics] {10.1088/0022-3700/20/18/023}, 20, 4869

\bibitem[\protect\citeauthoryear{OPEN-ADAS}{OPEN-ADAS}{2024}]{openadas}
OPEN-ADAS 2024, OPEN-ADAS website, \url{https://open.adas.ac.uk}

\bibitem[\protect\citeauthoryear{Parpia, {Froese Fischer}  \& Grant}{Parpia
  et~al.}{1996}]{grasp}
Parpia F.~A.,  {Froese Fischer} C.,   Grant I.~P.,  1996, Comp. Phys. Comm.,
  94, 249

\bibitem[\protect\citeauthoryear{Pian et~al.,}{Pian et~al.}{2017}]{pian2017}
Pian E.,  et~al., 2017, \mn@doi [Nature] {10.1038/nature24298}, 551, 67

\bibitem[\protect\citeauthoryear{Pognan, Grumer, Jerkstrand  \& Wanajo}{Pognan
  et~al.}{2023}]{Pognan2023}
Pognan Q.,  Grumer J.,  Jerkstrand A.,   Wanajo S.,  2023, \mn@doi [Monthly
  Notices of the Royal Astronomical Society] {10.1093/mnras/stad3106}, 526,
  5220

\bibitem[\protect\citeauthoryear{Reader \& Acquista}{Reader \&
  Acquista}{1997}]{Reader_1997}
Reader J.,  Acquista N.,  1997, \mn@doi [Physica Scripta]
  {10.1088/0031-8949/55/3/009}, 55, 310

\bibitem[\protect\citeauthoryear{Smartt et~al.,}{Smartt
  et~al.}{2017}]{smartt2017}
Smartt S.~J.,  et~al., 2017, \mn@doi [Nature] {10.1038/nature24303}, 551, 75

\bibitem[\protect\citeauthoryear{{Summers} et~al.,}{{Summers}
  et~al.}{2006}]{Summers2006}
{Summers} H.~P.,  et~al., 2006, \mn@doi [Plasma Physics and Controlled Fusion]
  {10.1088/0741-3335/48/2/007}, \href
  {https://ui.adsabs.harvard.edu/abs/2006PPCF...48..263S} {48, 263}

\bibitem[\protect\citeauthoryear{Vieira, Ruan, Haggard, Ford, Drout, Fernández
   \& Badnell}{Vieira et~al.}{2023}]{Vieira_2023}
Vieira N.,  Ruan J.~J.,  Haggard D.,  Ford N.,  Drout M.~R.,  Fernández R.,
  Badnell N.~R.,  2023, \mn@doi [The Astrophysical Journal]
  {10.3847/1538-4357/acae72}, 944, 123

\bibitem[\protect\citeauthoryear{{Vogl}, {Sim}, {Noebauer}, {Kerzendorf}  \&
  {Hillebrandt}}{{Vogl} et~al.}{2019}]{vogl_2019}
{Vogl} C.,  {Sim} S.~A.,  {Noebauer} U.~M.,  {Kerzendorf} W.~E.,
  {Hillebrandt} W.,  2019, \mn@doi [\aap] {10.1051/0004-6361/201833701}, \href
  {https://ui.adsabs.harvard.edu/abs/2019A&A...621A..29V} {621, A29}

\makeatother
\end{thebibliography}
\bsp
\label{lastpage}
\end{document}